\documentclass[final, authoryear, 5p,twocolumn,10pt]{elsarticle} 

\usepackage{amsmath,amssymb}
\usepackage{graphicx}
\usepackage{bm}
\usepackage[english]{babel}
\usepackage[usenames, dvipsnames]{xcolor}
\usepackage[normalem]{ulem}

\usepackage{lineno}

\newcommand{\bse}{\begin{subequations}}
\newcommand{\ese}{\end{subequations}}
\newcommand{\bfi}{\begin{figure}}
\newcommand{\efi}{\end{figure}}
\newcommand{\be}{\begin{equation}}
\newcommand{\ee}{\end{equation}}
\newcommand{\BE}{\begin{eqnarray}}
\newcommand{\EE}{\end{eqnarray}}
\newcommand{\dv}[2]{\frac{\mathrm{d} #1}{\mathrm{d} #2} }

\journal{Ecological Complexity} 

\begin{document}
\begin{frontmatter}

\title{Demographic noise and resilience in a semi-arid ecosystem model}
\author[uom,udc]{John Realpe-Gomez}
\ead{jrealpeg@unicartagena.edu.co}

\author[uu,cnr]{Mara Baudena}
\ead{m.baudena@uu.nl}

\author[uom]{Tobias Galla}
\ead{tobias.galla@manchester.ac.uk}

\author[uom]{Alan J. McKane}
\ead{alan.mckane@manchester.ac.uk}

\author[uu]{Max Rietkerk}
\ead{m.g.rietkerk@uu.nl}

\address[uom]{Theoretical Physics, School of Physics and Astronomy, 
The University of Manchester, Manchester M13 9PL, United Kingdom}

\address[udc]{Grupo de Ciencia Transdisciplinar, Informaci\'on, y Complejidad, Instituto de Matem\'aticas Aplicadas, Universidad de Cartagena, Bol\'ivar, Colombia }

\address[uu]{Department of Environmental Sciences, Copernicus Institute, 
Utrecht University, P.O. Box 80155 TC Utrecht, The Netherlands}

\address[cnr]{Grupo Interdisciplinar de Sistemas Complejos (GISC), Departamento de Matem\'aticas, Universidad Carlos III de Madrid, Avenida de la Universidad 30, 28911 Legan\'es, Madrid, Spain}

\date{\today}

\begin{abstract}
The scarcity of water characterising drylands forces vegetation to adopt appropriate survival strategies. Some of these generate water-vegetation feedback mechanisms that can lead to spatial self-organisation of vegetation, as it has been shown with models representing plants by a density of biomass, varying continuously in time and space. However, although plants are usually quite plastic they also display discrete qualities and stochastic behaviour. These features may give rise to demographic noise, which in certain cases can influence the qualitative dynamics of ecosystem models. In the present work we explore the effects of demographic noise on the resilience of a model semi-arid ecosystem. We introduce a spatial stochastic eco-hydrological hybrid model in which plants are modelled as discrete entities subject to stochastic dynamical rules, while the dynamics of surface and soil water are described by continuous variables.  The model has a deterministic approximation very similar to previous continuous models of arid and semi-arid ecosystems. By means of numerical simulations we show that demographic noise can have important effects on the extinction and recovery dynamics of the system. In particular we find that the stochastic model escapes extinction under a wide range of conditions for which the corresponding deterministic approximation predicts absorption into desert states. 
\end{abstract}

\end{frontmatter}


\section{Introduction} \label{s:i}

In arid and semi-arid ecosystems water constitutes the main limiting resource for vegetation. The harsh environmental conditions, related to frequent droughts, limit plant recruitment and survival and often prevent vegetation from fully covering the ground. In such areas vegetation largely occurs in patches of high density surrounded by bare soil, and forming spatial vegetation patterns \citep[e.g.][]{Aguiar1999,Barbier2006,Deblauwe2008}. These spatial structures emerge from the system's dynamics as a consequence of scale-dependent water-vegetation feedback mechanisms even in absence of any underlying spatial heterogeneity \citep[e.g.][]{Klausmeier1999,vonHardenberg2001,Rietkerk-2002,Meron2012}. For example, in arid areas the vegetation increases the infiltration capacity of the soil as compared to bare ground because of root penetration \citep{Rietkerk1997,Walker1981} and because it prevents the formation of biogenic crusts that would form in bare soil \citep{Casenave-1992}. While vegetation patches compete for water, vegetation enhances its own growth locally within the patches. This so-called infiltration feedback is known to be one of the most important scale-dependent water-vegetation feedback mechanisms, leading to self-organization and pattern formation phenomena \citep[e.g.][]{Rietkerk-2002}.

Scale-dependent resource concentration mechanisms of this type are also connected to the possible occurrence of catastrophic shifts in ecosystems. For example, a vegetated patchy state may turn into a degraded state with mostly bare soil if rainfall decreases below a certain threshold. A subsequent increase in rainfall above the threshold may not be enough to recover the previous vegetation state \citep[e.g.][]{vonHardenberg2001, Rietkerk2004,Rietkerk2008,Baudena2012b}. This is an example of how the resilience of such ecosystems is strongly interrelated with spatial structure. By `resilience' we mean here the ability of an ecosystem to remain in a given domain of attraction and to return quickly to the same state after disturbances \citep{Rietkerk2008}.

In models used to represent dryland ecosystems both water and plants are often represented as density fields, varying continuously in time and space, and
their dynamics is typically described by deterministic differential equations \citep[e.g.][]{Gilad2004,Rietkerk-2002}. In other models the plant dynamics is modelled using stochastic differential equations \citep{Manor2008}, or instead the vegetation is represented by discrete states in a drastically simplified way \citep[e.g.][]{Kefi2007,Kefi2007a}. 
Continuous models tend to be simple and provide powerful insight through the use of analytical techniques. Moreover, in the case of semi-arid ecosystems, they 
 very successfully point out the importance of scale-dependent water-vegetation feedbacks in the self-organised pattern formation processes \citep[e.g.][]{vonHardenberg2001,Rietkerk-2002}.

The spatial self-organization of drylands ecosystems has not yet been explored with models representing single plants individually, even though such individual-based models are now increasingly common in other areas of ecology and biology. 
A reason for this may be that plants are not always conceived as discrete individuals since, unlike animals, they are extremely plastic and can respond quite continuously to environmental changes, for instance by partially dying and recovering later on \citep{Crawley-1990}. However, plants have also discrete features: in their seed stage, they behave as discrete entities that can fall and perhaps germinate in some random position. The birth and establishment of a single plant is also subject to a collection of random events. Furthermore,
they can live alone in patches composed essentially of a single plant. Finally, the death of a whole plant is also a possible (unpredictable) event. These features can be readily accommodated in a stochastic individual-based modelling approach. To be more precise, plants in drylands often have a modular structure and are composed of multiple hydraulically independent stems \citep{Schenk-2008}. Except when the plant is in a seed stage such stems could be seen as the relevant individual entities rather than the whole plant. 
 In plant ecology, forest models are a successful example of individual-based modelling \citep{Botkin-1972, Shugart-1984, Pacala-1993, Pacala-1996, Moorcroft2001, Perry-2006}. The approach has been extended to grasslands as well \citep{Jorgensen-Handbook,Coffin-1990, Peters-2002, Rastetter-2003}, and it has also been used as a method to parametrise mean-field differential equations from fine-scale data \citep{Moorcroft2001}. In principle, it is possible to capture the most salient features of an IBM by a suitable continuous model \citep{McKane-TREE}. How to do this in general is an active area of research in the modelling of complex systems \citep{SanMiguel-2012}. In the case of forest models, good progress has been made in achieving this \citep{Pacala-2008}. Formally speaking, an IBM can coincide with a deterministic continuous model only in the limit of an infinite number of individuals. It might be imagined that a system with a relatively large number of individuals would simply lead to a small amount of noise around the deterministic solution. In some situations this is the case, but in others, however, the stochastic effects in an IBM can produce {qualitatively} different results \citep{McKane-PRL2005,Goldenfeld-2009,Biancalani-2010,Goldenfeld-2011,Biancalani-2011,Rogers-2012,Grima-Nature}. Still these effects become negligible with a `sufficiently' large number of individuals, but how large is `sufficiently' large? It can actually be  {unrealistically} large and the stochastic effects may turn out to be necessary to account for the observed behaviour of real systems. 
 It is not clear {\it a priori} how large the numbers need to be, and the system of interest needs to be carefully analysed before reaching a conclusion.

Following the previous discussion it is understandable that a deterministic continuous model can be a good approximation to an individual-based model of forests, where the vegetation is rather dense. In drylands, on the other hand, vegetation is composed of a relatively small number of plants coexisting with regions of empty land in which the above-ground biomass density is essentially zero.
Under these conditions the intrinsic stochastic behaviour of individual plants may turn out to be relevant for understanding the behaviour of the system \citep{McKane-TREE}. Even more so if the  drylands are  close to extinction, where the number of individuals is rather scarce.

In arid areas, vegetation is strictly dependent on water availability,
whose dynamics is effectively captured by a deterministic continuous approach. A natural approach to represent the dynamics of drylands is in terms of so-called `hybrid models', in which the dynamics of water is represented by differential equations, and in which the plant dynamics are described by an IBM  \citep{Vincenot2010}. Thus, these models combine both continuous and discrete variables.
 
In our work we use such a hybrid model to study the resilience of semi-arid ecosystems against desertification.  The model includes the water-vegetation infiltration feedback, and thus we expect vegetation patterns to emerge as previously observed in other types of models of semi-arid ecosystems \citep[e.g.][]{Rietkerk-2002}.  The use of individual-based models can be computationally demanding if too many details are included, and this can obstruct the identification of the underlying mechanisms relevant for the behaviour of the system. An example is the continuous range of values characterising each individual, e.g. mass and heterogeneity: every plant having a different mass.  Our approach constitutes a compromise between realism, tractability, and insight, concentrating on what we think could be the most relevant vegetation features. We extend an IBM, recently investigated by \cite{McKane-2012} to include spatial interactions. In particular, we neglect the growth phase and assume all plants have the same mass. We expect this to be a reasonable approximation if the time it takes for a plant to establish is much smaller than the time scale in which the spatial distribution of biomass changes appreciably and the growth phase of each individual plant is not very relevant for the system dynamics. We study the resilience and stability of the model dryland ecosystem when the individual nature of plants and its intrinsic stochasticity are considered. Our aim is to investigate the role of demographic noise and to understand whether noise enhances recovery or whether it may drive the system to extinction. To do so, we will compare the outcomes of a stochastic IBM and a deterministic continuous model.

\section{Model definitions}\label{s:gm}
\subsection{Stochastic model}
\subsubsection{General considerations}
\begin{figure*}
\centering
\begin{picture}(0,0)%
\includegraphics{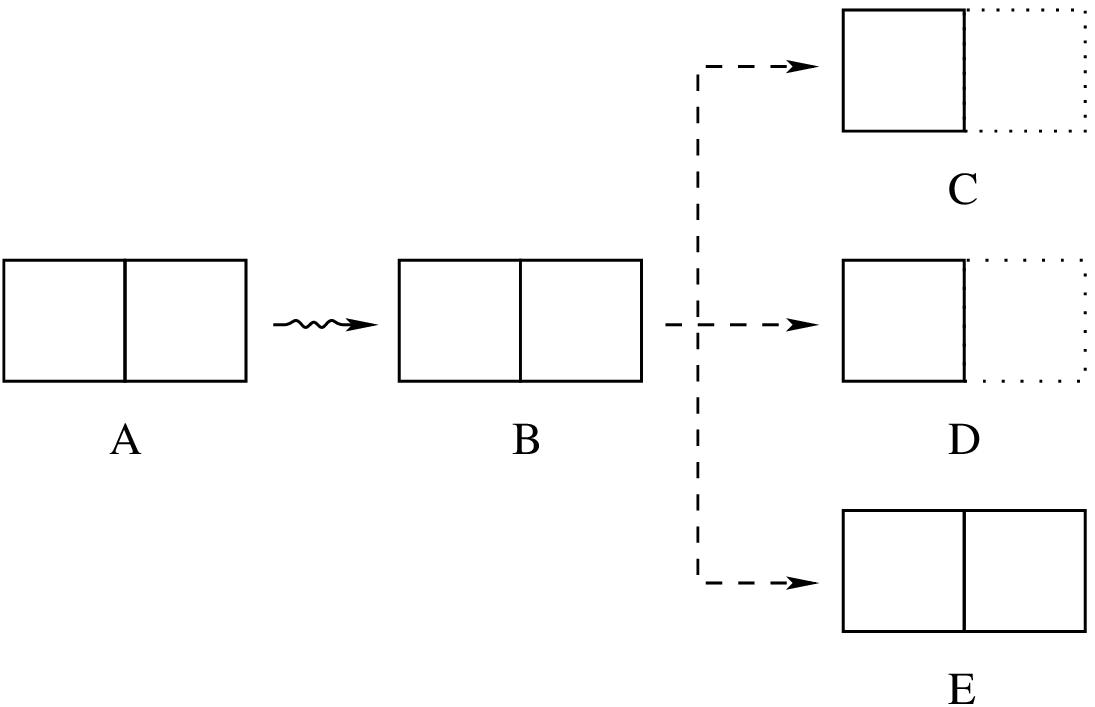}%
\end{picture}%
\setlength{\unitlength}{3398sp}%
\begingroup\makeatletter\ifx\SetFigFont\undefined%
\gdef\SetFigFont#1#2#3#4#5{%
  \reset@font\fontsize{#1}{#2pt}%
  \fontfamily{#3}\fontseries{#4}\fontshape{#5}%
  \selectfont}%
\fi\endgroup%
\begin{picture}(6074,4785)(-2046,-3946)
\put(-1754,-736){\makebox(0,0)[lb]{\smash{{\SetFigFont{11}{13.2}{\familydefault}{\mddefault}{\updefault}{\color[rgb]{0,0,0}Time $t$}%
}}}}
\put(406,-736){\makebox(0,0)[lb]{\smash{{\SetFigFont{11}{13.2}{\familydefault}{\mddefault}{\updefault}{\color[rgb]{0,0,0}Time $t+\tau$}%
}}}}
\put(-1619, 29){\makebox(0,0)[lb]{\smash{{\SetFigFont{9}{10.8}{\familydefault}{\mddefault}{\updefault}{\color[rgb]{0,0,0}DETERMINISTIC EVOLUTION}%
}}}}
\put(-1779,-196){\makebox(0,0)[lb]{\smash{{\SetFigFont{9}{10.8}{\familydefault}{\mddefault}{\updefault}{\color[rgb]{0,0,0}OF WATER BETWEEN $t$ AND $t+\tau $}%
}}}}
\put(1756,704){\makebox(0,0)[lb]{\smash{{\SetFigFont{9}{10.8}{\familydefault}{\mddefault}{\updefault}{\color[rgb]{0,0,0}STOCHASTIC BIRTH OR}%
}}}}
\put(1666,479){\makebox(0,0)[lb]{\smash{{\SetFigFont{9}{10.8}{\familydefault}{\mddefault}{\updefault}{\color[rgb]{0,0,0}DEATH OF PLANTS AT $t+\tau$}%
}}}}
\put(2701,-308){\makebox(0,0)[lb]{\smash{{\SetFigFont{12}{14.4}{\familydefault}{\mddefault}{\updefault}{\color[rgb]{0,0,0}${n_i+1}$}%
}}}}
\put(2701,-1703){\makebox(0,0)[lb]{\smash{{\SetFigFont{12}{14.4}{\familydefault}{\mddefault}{\updefault}{\color[rgb]{0,0,0}${n_i-1}$}%
}}}}
\put(2701,-601){\makebox(0,0)[lb]{\smash{{\SetFigFont{12}{14.4}{\familydefault}{\mddefault}{\updefault}{\color[rgb]{0,0,0}$\omega_i,\, \sigma_i$}%
}}}}
\put(2701,-1996){\makebox(0,0)[lb]{\smash{{\SetFigFont{12}{14.4}{\familydefault}{\mddefault}{\updefault}{\color[rgb]{0,0,0}$\omega_i,\, \sigma_i$}%
}}}}
\put(3376,-3098){\makebox(0,0)[lb]{\smash{{\SetFigFont{12}{14.4}{\familydefault}{\mddefault}{\updefault}{\color[rgb]{0,0,0}$n_j+1$}%
}}}}
\put(2701,-3391){\makebox(0,0)[lb]{\smash{{\SetFigFont{12}{14.4}{\familydefault}{\mddefault}{\updefault}{\color[rgb]{0,0,0}$\omega_i,\, \sigma_i$}%
}}}}
\put(3376,-3391){\makebox(0,0)[lb]{\smash{{\SetFigFont{12}{14.4}{\familydefault}{\mddefault}{\updefault}{\color[rgb]{0,0,0}$\omega_j,\, \sigma_j$}%
}}}}
\put(2881,-3098){\makebox(0,0)[lb]{\smash{{\SetFigFont{12}{14.4}{\familydefault}{\mddefault}{\updefault}{\color[rgb]{0,0,0}${n_i}$}%
}}}}
\put(-1079,-1703){\makebox(0,0)[lb]{\smash{{\SetFigFont{12}{14.4}{\familydefault}{\mddefault}{\updefault}{\color[rgb]{0,0,0}$n_j$}%
}}}}
\put(1081,-1703){\makebox(0,0)[lb]{\smash{{\SetFigFont{12}{14.4}{\familydefault}{\mddefault}{\updefault}{\color[rgb]{0,0,0}$n_j$}%
}}}}
\put(901,-1996){\makebox(0,0)[lb]{\smash{{\SetFigFont{12}{14.4}{\familydefault}{\mddefault}{\updefault}{\color[rgb]{0,0,0}$\textcolor{blue}{\omega_j},\, \sigma_j$}%
}}}}
\put(1981,-1636){\makebox(0,0)[lb]{\smash{{\SetFigFont{12}{14.4}{\familydefault}{\mddefault}{\updefault}{\color[rgb]{0,0,0}$\textcolor{Green}{n_i}\Gamma_d$}%
}}}}
\put(-1304,-1996){\makebox(0,0)[lb]{\smash{{\SetFigFont{12}{14.4}{\familydefault}{\mddefault}{\updefault}{\color[rgb]{0,0,0}${\omega_j^\prime},\, \sigma_j^\prime$}%
}}}}
\put(-1956,-1996){\makebox(0,0)[lb]{\smash{{\SetFigFont{12}{14.4}{\familydefault}{\mddefault}{\updefault}{\color[rgb]{0,0,0}$\omega_i^\prime,\, \sigma_i^\prime$}%
}}}}
\put(1801,-3796){\makebox(0,0)[lb]{\smash{{\SetFigFont{12}{14.4}{\familydefault}{\mddefault}{\updefault}{\color[rgb]{0,0,0}$\textcolor{Green}{n_i}\Gamma_s(\textcolor{blue}{\omega_j})$}%
}}}}
\put(226,-1996){\makebox(0,0)[lb]{\smash{{\SetFigFont{12}{14.4}{\familydefault}{\mddefault}{\updefault}{\color[rgb]{0,0,0}$\textcolor{red}{\omega_i},\, \sigma_i$}%
}}}}
\put(406,-1703){\makebox(0,0)[lb]{\smash{{\SetFigFont{12}{14.4}{\familydefault}{\mddefault}{\updefault}{\color[rgb]{0,0,0}$\textcolor{Green}{n_i}$}%
}}}}
\put(-1799,-1703){\makebox(0,0)[lb]{\smash{{\SetFigFont{12}{14.4}{\familydefault}{\mddefault}{\updefault}{\color[rgb]{0,0,0}${n_i}$}%
}}}}
\put(1846, 74){\makebox(0,0)[lb]{\smash{{\SetFigFont{12}{14.4}{\familydefault}{\mddefault}{\updefault}{\color[rgb]{0,0,0}$\textcolor{Green}{n_i}\Gamma_b(\textcolor{red}{\omega_i})$}%
}}}}
\put(-809,-1186){\makebox(0,0)[lb]{\smash{{\SetFigFont{11}{13.2}{\familydefault}{\mddefault}{\updefault}{\color[rgb]{0,0,0}Eqs. (\ref{e:model})-(\ref{e:sat})}%
}}}}
\end{picture}%
\caption{(Colour online) Illustration of the stochastic hybrid model for semi-arid ecosystems. For clarity we show only two neighbouring cells $i$ and $j$ but the same applies to all the other cells. Suppose that, at time $t$, the system is in a state where the number of plants and the depths of soil and surface water in cell $i$ are given by $n_i$, $\omega_i^\prime$ and $\sigma_i^\prime$, respectively (A). The analogous quantities in cell $j$ are $n_j, \omega_j', \sigma_j'$. Suppose, furthermore, that the next birth or death of a plant takes place at time $t+\tau$;  these events happen at random, and so $\tau$ is itself a stochastic variable. Since there are no transition events between $t \text{ and } t+\tau$, soil and surface water in all cells evolve deterministically according to Eqs. (\ref{e:model})-(\ref{e:sat}) in this time interval. Suppose now that their new state at $t+\tau$ is given by $\omega_i$ and $\sigma_i$, for all cells $i$ (B). At $t+\tau$ a stochastic transition happens, there are three possible types of transitions: a plant at a cell $i$ gives birth to a new plant  in the same cell ({C}), it dies (D) or it gives birth to a new plant in a neighbouring cell $j$ (E). These transitions happen with rates $n_i \Gamma_b(\omega_i)$, $n_i \Gamma_d$, and $n_i \Gamma_s(\omega_j)$, respectively (see Eqs. (\ref{e:rates}) and (\ref{e:transition})). Immediately after $t+\tau$, and until the next birth or death of a plant takes place, soil and surface water in all cells again evolve deterministically as before.}\label{f:model}
\end{figure*}

In this section we introduce our stochastic hybrid model for semi-arid ecosystems. The model describes the dynamics of vegetation, soil and surface water in a given area to which the model is applied. The source of surface water is rainfall. Surface water then infiltrates the soil where plants can take it up. Plants are considered as discrete entities following a stochastic dynamics \citep{McKane-2012}, while soil and surface water are modelled by continuous variables whose dynamics would be deterministic \citep{Pueyo2008, Rietkerk-2002, Rietkerk-2001} were they not influenced by plants. Figure \ref{f:model} summarises the dynamics of the model. For computational simplicity we will work mostly in one spatial dimension. Occasionally we will also consider the case of two dimensions; this is to illustrate that the results of the one-dimensional model are still valid in this more realistic case. We will define the model in one spatial dimension only, the modifications required to extend it to two dimensions are straightforward. In the one-dimensional model we assume that the land is partitioned into $L$ uniform square cells, each labelled by an index $i$. Thus the model describes a linear array of square cells. The length of the side of a cell is denoted by $h$.  With every cell $i$ we associate three variables corresponding to the discrete (integer) number of plants, $n_i$, in the cell and the continuous quantities of soil and surface water, $\omega_i$ and  $\sigma_i$, respectively, on that piece of land. We will assume that the density of biomass per unit area in cell $i$ is given by $\rho_i=\mu \, n_i$, where $\mu$ represents the mass of one plant individual divided by the area of the cell, i.e. $\mu=m_P/h^2$. We have made here the simplifying assumption that all plants have identical mass $m_P$. Thus, $\mu$  characterises the contribution of an individual plant to the biomass density. We consider the cells as homogeneous, i.e. we do not take into account any structure within the cell, such as the spatial distribution of plants within a cell. We also neglect properties of individual plants, such as for example different plant sizes or stages of growth. The detailed dynamics of the various components of the model are explained below, where we also introduce the relevant model parameters. The units and numerical values of all the parameters in the model are summarised in Table \ref{t}.

\subsubsection{Water dynamics}
For all cells, $i$, the dynamics of the depths of soil and surface water, $\omega_i$ and $\sigma_i$ (measured in mm) are described by differential equations that are deterministic when the biomass density $\rho_i$ (measured in g m$^{-2}$) is kept constant; namely

\begin{subequations}\label{e:model}
\BE
\dv{\omega_i}{t} &=& {F_\omega}(\rho_i, \omega_i ,\sigma_i) + {D_\omega}\, \Delta\, \omega_i\, ,\\
\dv{\sigma_i}{t} &=& {F_\sigma}(\rho_i, \omega_i ,\sigma_i) + {D_\sigma}\, \Delta\, \sigma_i\, .
\EE
\end{subequations}
In each of these equations the first term on the right-hand side describes the water dynamics within a cell. These are captured by the functions $F_\omega$ and $F_\sigma$, to be specified below (see Eqs.~(\ref{e:ns}) and (\ref{e:sat})). The second term in each of the two equations describes spatial transport processes{, that we model here as standard diffusive processes by using the discrete diffusion operator}

\begin{subequations}\label{e:diff}
\BE
\Delta {\omega}_i &=& \frac{1}{h^2}\sum_{j\in N(i)}(\omega_j -\omega_i)\, ,\\
\Delta {\sigma}_i &=& \frac{1}{h^2}\sum_{j\in N(i)}(\sigma_j -\sigma_i)\, .
\EE
\end{subequations}
Here the sum is over all elements $j\in N(i)$, i.e. the set of cells $j$ which are nearest neighbours of cell $i$. The coefficients $D_\omega$ and $D_\sigma$ of the diffusion operator in Eqs. (\ref{e:model}) are the diffusion constants corresponding to the transport of soil and surface water, respectively. {This is usually a good approximation \citep[see e.g.][]{Rietkerk-2002} of the more accurate description derived from shallow water theory \citep{Gilad2004,Meron-2011}.} {Indeed, recent work \citep{Stelt-2013} suggests that there is no qualitative difference between these two descriptions.} 

Following \cite{Rietkerk-2001} we define the functions $F_\omega$ and $F_\sigma$, describing the on-site water dynamics, as
\begin{subequations}\label{e:ns}
\begin{align}
F_\omega (\rho, \omega, \sigma) & = \alpha(\rho)\, \sigma \, -\, \beta(\omega)\, \rho\, -\, r\, \omega\, , \label{se:ns_soil}\\
F_\sigma (\rho, \omega, \sigma) & = R\, -\, \alpha(\rho)\,  \sigma \, . \label{se:ns_surface}
\end{align}
\end{subequations}
Here,
\be\label{e:sat}
\alpha(\rho) = a\: \frac{\rho +k_2W_0}{\rho +k_2},\hspace{0.5cm}
\beta(\omega) = b\:\frac{\omega}{\omega + k_1}, 
\ee
describe the infiltration rate of surface water into the soil, and the soil water uptake due to plants, respectively. These rates saturate for large values of $\rho$ and $\omega$, respectively. The model parameters $k_1$ and $k_2$ determine the exact shape of the saturation curves and are known as `half-saturation' constants \citep{Rietkerk-2001}. The dependence of $\alpha$ on the biomass density, $\rho$, introduces the infiltration-water-vegetation feedback, known to generate spatial vegetation patterns in existing deterministic models \citep[e.g.][]{Rietkerk-2002}. The parameter $r$ characterises the loss of soil water due to evaporation (Eq. (\ref{e:ns}a)), while the parameter $R$ in Eq.~(\ref{e:ns}b) is the rainfall rate. The dynamics of water are illustrated in Fig. \ref{f:model}, see the left-most pair of cells (A) and the pair of cells in the middle (B).

\subsubsection{Plant dynamics}
We model individual plants with an IBM. Individual plants $I_i$ in cell $i$ are represented as discrete entities whose dynamics are given by the stochastic transition rules

\bse\label{e:rules}
\begin{align}
I_i &\xrightarrow{\Gamma_b(\omega_i)} 2 I_i\, \label{se:dispersal0},\\
I_i &\xrightarrow{\Gamma_d} E_i, \label{se:dispersal1}\\
I_i &\xrightarrow{\Gamma_s(\omega_j)}  I_i + I_j,\hspace{1cm} j\in N(i)\, .\label{se:dispersal}
\end{align}
\ese
Here the symbols over the arrows refer to the probabilities per unit time, or rates, of the corresponding transitions. These rates may depend on the amount of available soil water $\omega$ (see Eqs.~(\ref{e:rates}) below). The first transition rule,  Eq.~(\ref{se:dispersal0}), corresponds to an individual plant $I_i$ in cell $i$ giving birth to another plant in the same cell $i$ at a rate $\Gamma_b(\omega_i)$. The second rule, Eq.~(\ref{se:dispersal1}), refers to the death of a plant in cell $i$. The quantity $E_i$ on the right-hand side of this reaction stands for a vacancy in cell $i$, so that this transition rule indicates that an individual plant $I_i$ in a cell $i$ dies at a rate $\Gamma_d$ and leaves behind an empty place $E_i$ in cell $i$. The third rule, Eq.~(\ref{se:dispersal}), describes a process in which an individual plant in cell $i$ gives birth to another plant in a neighbouring cell $j$ at a rate $\Gamma_s(\omega_j)$. This captures the dispersal of the seeds of an individual plant to neighbouring cells and includes the probability of germination of the seedling. Notice that the rate of such dispersion processes depends on the amount of soil water, $\omega_j$, in the cell in which the new plant is born (see Eq. (\ref{e:rates}) below). 

We define the transition rates in analogy with previous deterministic models, e.g. \cite{Rietkerk-2001}, \cite{Rietkerk-2002}, and \cite{Pueyo2008}. Specifically, we use
\bse\label{e:rates}
\begin{align}
\Gamma_b(\omega_i) &= \widetilde{c}\: \beta(\omega_i),\\
\Gamma_d &= d \, , \\
\Gamma_s(\omega_j) &=  K\: \widetilde{c}\: \beta(\omega_j)\, .
\end{align}
\ese
The plant birth rate $\Gamma_b$ is proportional to the plant water uptake $\beta$ (Eq. (\ref{e:sat})), and $\widetilde{c}$ is a constant that characterises the conversion rate at which water uptake is turned into newly established plants (Eq. (\ref{e:rates}a)). 
The death rate $\Gamma_d$  is assumed to be a constant $d$ (Eq. (\ref{e:rates}b)). The rate $\Gamma_s$ with which new plants are established in neighbouring cells is proportional to the water uptake $\beta$, to the parameter $\widetilde{c}$  and to the constant $K$, representing the probability that a seed falls into a neighbouring cell and manages to survive (Eq. (\ref{e:rates}c)). As described above the water uptake rate $\beta$, and thus $\Gamma_{s}$, depend on the soil water concentration in the colonised cell ($\omega_j$). This reflects the fact that the probability of germination of the seedling in the neighbouring cells is a function of the local availability of soil water. In our simple model we assume that plants can colonise only nearest neighbour cells \citep[see e.g.][]{Kefi2007a}. This can be generalised to take into account longer-range dispersal kernels, which may be appropriate when the characteristic scale of seed dispersal is appreciably greater than $h$, the lateral extension of a cell \citep{Pueyo2008}. It is worth pointing out that the transition rates do not depend on the whole history of the system, but only on its current state. This type of stochastic hybrid model is usually referred to as a {piecewise deterministic} Markov process in the literature \citep{Davis-1984, Faggionato-2009, Faggionato-2010, McKane-2012}.
 
\subsubsection{Coarse-graining: cell dynamics}

Given that we neglect the spatial structure within a cell, the relevant transition rules involve the state of cells rather than that of particular individual plants, so our model is of an effective nature; it provides a coarse grained description of the underlying dynamics. According to the rules for individual plants, Eq. (\ref{e:rules}), the only possible transitions for a cell $i$ with $n_i$ plants are $n_i\to n_i\pm 1$, i.e. the birth or death of a plant in cell $i$. Hence, the possible transitions that can occur in cell $i$ are fully specified by the expressions  (see e.g. \cite{McKane-TREE} for similar models)
\bse\label{e:transition}
\begin{align}
{T}_b(n_i+1|n_i;\, \omega_i) &= n_i\, \Gamma_b(\omega_i), \\
{T}_d(n_i-1|n_i) &= n_i\, \Gamma_d, \\
{T}_s^{i\to j}(n_i, n_j+1|n_i, n_j;\, \omega_j) &= n_i\, \Gamma_s(\omega_j), \hspace{0.5cm} j\in N(i) ,
\end{align}
\ese
for all $i$. The notation $T_b(n_i +1|n_i ;\omega_i)$, for instance, stands for the probability per unit time (or rate) that the number of plants $n_i$ in a cell $i$ increases by one, given that the amount of soil water in $i$ is $\omega_i$ at the time the transition takes place. Since each plant within a cell $i$ has the potential to give birth to a new plant in $i$ with a rate $\Gamma_b(\omega_i)$ independently of all other plants, the total rate for a transition $n_i\to n_i +1$ in cell $i$ is given by $n_i\, \Gamma_b(\omega_i)$. Similar considerations apply for the other two processes described in Eq. (\ref{e:transition}) above. Notice that the arguments before the vertical bar in the above rate functions, Eqs. (\ref{e:transition}), refer to the state of the system after the transition, while those to the right of the vertical bar refer to the state before the transition, following the standard notation for conditional probabilities. This stochastic dynamics of cells is illustrated in Fig. \ref{f:model}, see the  pair of cells in the middle (B) and the three right-most pairs of cells (C, D, E). The cells at the top (C) correspond to an on-site birth event, those in the middle (D) correspond to a death event and those at the bottom (E) correspond to a birth event induced by a neighbouring cell.

\subsection{Deterministic approximation}\label{s:dl}
\subsubsection{General considerations}
The dynamics of plants in our model are stochastic, so we cannot predict the exact state of the system at a future time. At best, it might be possible to obtain the probability of finding the system in a given state, specified by the number of plants, $n_i$, and the soil and surface water depths $\omega_i$ and $\sigma_i$, in all cells $i$.  From such probability distributions it would then be possible to compute the expected average behaviour. Another way to study stochastic systems is to perform a certain number $S$ of independent stochastic simulations.  In doing so we could, for each cell $i$ and for a given time $t$, estimate the average density of biomass or the average depths of soil and surface water. We will denote these quantities by $P_i$, $W_i$ and $O_i$, respectively. 

The transition rules governing the dynamics of plants (Eqs. (\ref{e:rates})-(\ref{e:transition})) along with the dynamical laws for the soil and surface water (Eqs. (\ref{e:model})-(\ref{e:sat})) induce a corresponding deterministic equation for the dynamics of these averages in the limit of a very large number of simulations, formally $S\to \infty$. At the same time this deterministic approximation also describes the behaviour of the stochastic model for very small $\mu$. The equivalence is formally exact in the limit $\mu\to 0$ \citep[cf.][]{McKane-2012}. In this sense $\mu$ controls the strength of the noise in the stochastic model: if $\mu\to 0$ the noise is irrelevant and the dynamics becomes deterministic. Since $\rho_i = \mu n_i$ the number of plants tends to be very large ($n_i\to \infty$) in this limit in regions with a non-zero density of biomass, $\rho_i>0$. We also refer to \citet{McKane-2012} for more details of a non-spatial version of the model; see also \citet{McKane-TREE} for a general review of individual-based models and the connection with deterministic limiting descriptions.
 
The deterministic model governing the average behaviour of the stochastic model turns out to be similar to models based on partial differential equations, well known in the literature \citep{Rietkerk-2001, Rietkerk-2002, Pueyo2008, Rietkerk-2012}. The average biomass density, $P_i$, and the average depths of soil and surface water, $W_i$ and $O_i$, respectively, follow the equations
\begin{subequations}\label{e:pde}
\BE
\dv{P_i}{t} &=& F_\rho(P_i,W_i, O_i) +  D_\rho\left(W_i\right)\, \Delta P_i,\label{e:pdea}
\\
\dv{W_i}{t} &=& {F_\omega}(P_i, W_i ,O_i) + {D_\omega}\, \Delta\, W_i\, ,\label{e:pdeb}\\
\dv{O_i}{t} &=& {F_\sigma}(P_i, W_i ,O_i) + {D_\sigma}\, \Delta\, O_i\, .\label{e:pdec}
\EE
\end{subequations}
Here $F_\omega$ and $F_\sigma$ are given by Eqs. (\ref{e:ns})-(\ref{e:sat}) and
\be\label{e:Frho}
F_\rho(P,W,O) = c\beta\left(W\right)P - d P ,
\ee
where we have defined a rescaled conversion rate ${c} = (1 + z\, K)\, \widetilde{c} $, with $z$ the number of nearest neighbours of a cell, e.g. $z=2$ in one dimension and $z=4$ for two dimensions. The diffusion operator is given by Eq. (\ref{e:diff}) as before, and so Eq. (\ref{e:pdea}) above is effectively a diffusion equation with a diffusion coefficient $D_\rho(W_i) = h^2\, K\, \widetilde{c}\, \beta\left(W_i\right)$. The biomass diffusion coefficient thus depends on the average amount of soil water in the cell under consideration. {The reason for this water dependence is that the `diffusion' of biomass occurs when a plant in a given cell $j$ produces a newly established plant in a neighbouring cell ($i\in\partial j$). The assumption underlying our model is that the establishment of a new plant depends on the amount of water available in the cell in which the new plant is established. Indeed, a water-dependent diffusion coefficient is also implicit in the work by \cite{Pueyo2008}. In the special case in which the range of the dispersal kernel is so short that only dispersal to neighbouring cells occurs, Eqs. (\ref{e:pde}) above coincide with the equations obtained in \cite{Pueyo2008}. On the other hand, such a water dependence of the biomass `diffusion' coefficient constitutes the only difference between the deterministic equations  (\ref{e:pde}), and a discretisation of the set of equations introduced by \cite{Rietkerk-2001}.}

\subsubsection{Homogeneous fixed points}

The deterministic equations introduced above, Eq. (\ref{e:pde}), have homogeneous fixed points, i.e.~equilibrium states, such that $P_i = P$, $W_i = W$ and $O_i=O$, for all $i$, and
\be\label{e:nsfp}
\dv{P}{t}  =  0,\hspace{0.5cm}\dv{W}{t}  =  0,\hspace{0.5cm}\dv{O}{t}  =  0.
\ee
Such fixed points may or may not be stable under small spatially homogeneous perturbations. Depending on the choice of parameters there can be either one out of two possible stable fixed points \citep{Pueyo2008} or none \citep{McKane-2012}. The two types of stable fixed points correspond to a bare soil state (also referred to as a `desert state' in the following) 
\be\label{e:dfp}
P_d = 0,~W_d = \frac{R}{r},~O_d = \frac{R}{a\, W_0},
\ee 
or to a state with non-zero vegetation 
\be\label{e:vfp}
P_v = \frac{c\, R}{d}-\frac{r\, c\, k_1}{c\, b-d},~ W_v = \frac{d\, k_1}{c\, b-d},~ O_v = \frac{R}{\alpha(P_v)}.
\ee 
In the region where there are no stable fixed points, a numerical integration of the corresponding non-spatial equations shows the existence of homogeneous limit cycles{, as it has been reported recently by \cite{McKane-2012}. These oscillations become unstable once spatial structure is taken into account {giving way to vegetation patterns as discussed below}. In this work we will {not} focus on this parameter regime {since we are interested in a regime where the deterministic dynamics can lead to extinction (see below).}} 

\subsubsection{Spatial patterns and resilience}
Besides the homogeneous states given by Eqs. (\ref{e:dfp})-(\ref{e:vfp}) above, the deterministic system defined by Eqs. (\ref{e:pde}) can also display stationary spatial patterns \citep{Rietkerk-2002}. This is not surprising, given the similarity of Eqs. (\ref{e:pde}) to those investigated by \cite{Rietkerk-2001} and \cite{Pueyo2008}, which are known to display spatially patterned solutions. For certain values of the parameters, these patterns can `emerge' out of the homogeneous vegetated state given by Eq. (\ref{e:vfp}) when a small {heterogeneous perturbation is applied \citep{Rietkerk-2002}. This results in Turing patterns \citep{Turing-1952, Cross-book} and can be studied mathematically within a linear approximation of Eqs. (\ref{e:pde}) around the fixed point with homogeneous vegetation. {However, numerical integration of the full set of nonlinear equations has revealed that spatial patterns exist in a region of parameters much wider than that explained by such a linear analysis \citep{Rietkerk-2002}. These non-trivial solutions are inherently due to nonlinearities and so they cannot be captured in a linear approximation.} In a parameter range relevant for dryland ecology (i.e. at low rainfall values) a relatively large region of bistability between spatial patterns and the desert state exists. {In the region we investigate here the deterministic system can converge either to the desert state given by Eq. (\ref{e:dfp}) or to a stable vegetation pattern, depending on the initial conditions from which the dynamics are started.}  The basin of attraction of the state with vegetation patterns, i.e. the set of all initial conditions which lead to such a state, can be used to characterise the resilience of the deterministic model: the larger the size of the basin of attraction of a `desirable' state the more resilient is the system \citep{Holling-1973,Grimm-2011,Martin-2011}. Performing such a study quantitatively is not an easy task given that the space of all possible initial conditions is huge. We will therefore restrict our work to an analysis involving only one type of realistic initial conditions. This will be discussed in more detail below.

\section{Analysis}\label{s:analysis}

All results discussed in this paper were obtained from spatially explicit numerical simulations, mostly in one dimension, with a few two-dimensional examples for illustrative purposes. We used periodic boundary conditions throughout. In order to obtain sample simulations for the stochastic model we used an adaptation of an algorithm due to \cite{Gillespie-1976} \citep[see also][]{Gillespie-1977}, details are discussed in \citet{McKane-2012}. To obtain solutions of the deterministic approximation, Eq. (\ref{e:pde}), we integrated these equations numerically using the Euler-forward method with a time step of $\Delta t = 0.01$ days. 

In all cases we used the same initial conditions for both the stochastic model and its deterministic approximation obtained as follows: (a) soil water $\omega_i$, and surface water $\sigma_i$ in all cells $i$ were assigned the values corresponding to the homogeneous desert state given by $W_d$ and $O_d$ in Eq. (\ref{e:dfp}); (b) biomass density was assigned a value $\rho_0> 0$ in a fraction $f$ of the cells picked at random, and a value of zero in the remaining fraction $(1-f)$ of cells. We emphasise that in each run the random choice of populated cells is the same in both models. Notice that $\rho_0$, along with the parameter $\mu$, fixes the initial number of individual plants per initial populated cell $n_0 = \rho_0/\mu$, which has to be an integer. The possible choices of $\rho_0$ and $\mu$ are therefore constrained. 

{Before continuing we would like to comment on why we have chosen this particular type of initial conditions here. The main reason for this is that we are dealing with spatial models and to define a generic initial condition we would need a very large number of parameters, i.e. initial biomass, soil and surface water in each cell.} {It is to avoid this that we decided to focus here on a subset of initial conditions that can be easily specified by a few parameters, as we have explained above.} 

First we analysed the differences between the deterministic and the stochastic approach  comparing the behaviour of the system in single runs. Then, in order to study how generic the observed outcomes were, we estimated the probability of extinction, $P_{\rm ext}$, in both the stochastic and the deterministic models as a function of the different model parameters. This probability was obtained from simulations of the corresponding dynamics for several different initial conditions, picked at random as described above, and counting the fraction of samples in which the systems became extinct. In the case of the deterministic system the only difference between simulation runs is the randomly chosen initial spatial distribution of the vegetated cells. Once the initial condition is fixed, the subsequent dynamics is fully deterministic, with no further random elements. Similarly, the simulation runs of the stochastic system also involved a random element due to the initial conditions, but further stochasticity then entered during the actual run due to the random processes determining the sequence and timing of the transitions of the individual-based model.

We estimated the probability of extinction in terms of what we expect to be the relevant parameters of the set of initial conditions:  (a) the initial cover $f$, i.e. the fraction of cells with $\rho_0 >0$ in the initial conditions; (b) the initial amount of biomass $\rho_0$ introduced in those cells; (c) the mass of a plant per unit area $\mu$; and (d) the rainfall rate $R$. We chose $f$ and $\rho_{0}$ as relevant parameters because they determine the amount of initial biomass, $\mu$  because it controls the strength of the stochasticity in the model and $R$ in order to investigate the models under different conditions of water availability. In the limit $\mu\to 0$, the stochastic results are expected to show similar behaviour to the deterministic approximation, while for larger values of $\mu$ we would expect stochastic effects to be more common. 

To estimate the probabilities of extinction, $50$ simulations were run for each model and for each set of parameters. Each of these simulations was run for a period of time of up to $T=5000$ days. Simulation runs of the stochastic model were identified as becoming extinct if all cells $i$ reached a state with $n_i = 0$ within this time. In the deterministic model we applied a threshold criterion $P_i < \epsilon$, i.e. a given cell is assumed to contain no plants when its plant density is below a set threshold $\varepsilon$.  In order to be consistent with the original stochastic model we chose $\epsilon = \mu$, as we assumed that $\mu$ was the contribution of a single plant to the biomass density. The introduction of a threshold was necessary because the deterministic approximation works with a genuinely continuous density, and the state with exactly zero biomass is reached only asymptotically at infinite times. We carried out some consistency checks, and observe that results depend little on the exact choice of the value of the threshold. Findings remained essentially the same even for $\epsilon$ much smaller than $\mu$.
\section{Results}
\subsection{Dynamics}

We compared the dynamics of the stochastic model with the dynamics of the corresponding deterministic approximation. Both models were initialised with homogeneous initial conditions (i.e. $f=1$) with $\rho_0=10$ g m$^{-2}$. For all cells, soil and surface water took values corresponding to the desert state, given by Eq. (\ref{e:dfp}). We compared the time dynamics of the total biomass density for both models. As expected from the stability analysis of the deterministic non-spatial model, we observed that in this case the deterministic approximation converges asymptotically to the desert state (Fig. \ref{f:homo}, red thick line). In contrast the stochastic model did not reach extinction for the same homogeneous initial conditions and parameter values. The stochastic model initially followed a path close to the deterministic approximation and approached extinction, but then some regions displayed an explosive growth and the system finally was found in a quasi-stationary pattern. As expected the stochastic dynamics deviated substantially from its deterministic approximation especially when the number of plants in the system was small (see Fig. \ref{f:homo}). We will refer to this effect of escaping a path to extinction as `self-recovery'. From a mathematical perspective this phenomenon appears similar to the one investigated in \citet{McKane-WKB-2011}. For clarity the figure shows the results up to time $t=500$ days even though we ran the simulations for much longer (up to time $T=5\times 10^4$ days). We observed that the total biomass in the stochastic model remains essentially the same up to stochastic deviations, i.e. it reaches a stationary state. Still there was a nonzero probability that the stochastic model, after having escaped the initial path to extinction, e.g. after, say, $t\approx 300$ days in Fig. \ref{f:homo}, reached extinction. Such events occur because of large stochastic deviations, and the probability of such events can be expected to be rather small \citep{Frey-2010}. 

The phenomenon of self-recovery in the stochastic model can also be seen in a comparison of the time-dependent spatial  biomass density profile along the line of cells in both models, see Fig. \ref{f:patternzoom}. Initial conditions for the stochastic and the deterministic models were chosen as $f=1/2$ and $\rho_0 = 10$ g m$^{-2}$. For all cells the soil and surface water densities were initialised from the values of the desert state given by Eq. (\ref{e:dfp}). In Fig. \ref{f:patternzoom} we compare the dynamics of the vegetation profile in the stochastic model with that of the corresponding deterministic approximation. Again, we can see that the deterministic approximation leads to extinction while the stochastic model recovers. If we look at the stochastic simulation for a longer time  ($5\times 10^4$ days) we clearly see that vegetation persists, see Fig. \ref{f:pattern} below the horizontal dashed line. We have also used the last configuration reached by the stochastic model (horizontal dashed line in Fig. \ref{f:pattern}) as initial conditions for the deterministic model; the corresponding time dynamics is shown in Fig. \ref{f:pattern}, above the horizontal dashed line. The patterns also remain stable under the deterministic dynamics, but it was demographic stochasticity which allowed the system to escape extinction in the first place; running the deterministic dynamics alone leads to extinction. The stochastic system thus generated suitable initial conditions for the deterministic system to converge to a spatial pattern. In Fig. \ref{f:pattern} one can also observe that the patches reached by the deterministic model are fully frozen at long times while those generated by the stochastic model remain dynamic, they can split, merge or become extinct.

Similar observations can also be made in the two-dimensional system, as we show in Fig. \ref{f:pattern2d} and in the enclosed supplementary video animation. As expected, both the stochastic and the deterministic models display spatial vegetation patterns, similar to those observed previously with other models, and in an analogous range of rainfall values \citep{Rietkerk-2001,Rietkerk-2002,Pueyo2008}.

Finally we notice that in many simulations the stochastic system follows a regime of `explosive' {local} growth soon after it has escaped the path to extinction. During this relatively short period of time some cells can get to carry a relatively large number of individuals with a maximum of about $120$ plants observed in simulations. Afterwards plants spread out in space to a certain extent, and eventually the system appears to reach a stochastic stationary state (see supplementary video).

\subsection{Probability of extinction}

Next we will present the results of a more systematic investigation of the effect of `self-recovery'. Figure \ref{f:probvsf} shows the probability of extinction $P_{\rm ext}$ as a function of $f$, for two different values of $\rho_0$ ($10$  g m$^{-2}$ and $~50$ g m$^{-2}$). We observe that the stochastic model almost never reached extinction, while the deterministic approximation did so in almost all samples with $f>0.4$. These observations may appear counter-intuitive at first sight: in the deterministic model we find that the more plants are in the system initially, i.e. the larger the initial cover $f$, the more likely it is that the system reaches extinction. However, this effect is already seen in the deterministic non-spatial model which predicts the extinction of homogeneous initial configurations (i.e. $f=1$). The spatial model instead predicts stable patterns \citep{Rietkerk-2001, Rietkerk-2002, Pueyo2008}. Even more surprisingly perhaps, is that the larger the initial amount of biomass in each populated cell, $\rho_0$, the larger the probability of extinction tends to be. {To be more precise, this only happens if the initial cover is not too small ($f\gtrsim 0.1$)}. From the simulations, it seems that if the amount of biomass in a cell is too large, a fast spatial spread of plants is triggered, which favours a more homogeneous distribution of biomass. This amounts to an effective increase of the initial cover, $f$, and thus the previous effect takes over. {This behaviour appears to run contrary to the infiltration feedback mechanism for pattern formation: local biomass increase favours infiltration in detriment of water availability in the surroundings. However, a similar effect already exists in the deterministic model as well. Indeed, { in the region of} parameter values that we considered here (see Tab. \ref{t}), vegetation patterns are {not} the only stable stationary solutions to the deterministic system: homogeneous desert is also one such solutions; in other words, there is bistability between vegetation patterns and desert. This suggests that, in this regime, the infiltration feedback mechanism is not always effective in inducing the formation of patterns. To develop some intuition of the kind of processes that could lead to these phenomena, it is useful to imagine an {extreme case with relatively} large total reproduction rate and a {not so large average} transport of water. We can then expect that plants manage to spread quickly before water heterogeneities consolidate. This might lead to a rather homogeneous state with a behaviour similar to what one would see in the non-spatial model in this regime: all plants die together. Although this situation may not be realistic, it indicates that the infiltration feedback mechanism may break down at some point, leading the system to a desert state rather than to pattern formation. This appears to be consistent with our simulations, in which we find that a larger initial cover or a larger value of $\rho_0$ imply a larger total reproduction rate.}

We study the variations of the probability of extinction $P_{\rm ext}$ as a function of $\mu$ for three different values of $f$: $1/8$, $1/2$, and $7/8$ in Fig. \ref{f:probvsmp}. We can observe that $P_{\rm ext} \approx 0$ for almost all values of $\mu$ except for $\mu = 0.1$ g m$^{-2}$ and $\mu > 6$ g m$^{-2}$, which correspond to very small and large plant biomass, respectively.   

Next, we study the probability of extinction $P_{\rm ext}$ as a function of the rainfall rate $R$, for a range corresponding to typical dryland values, using three different values for $\mu$: $1$ g m$^{-2}$, $~5$ g m$^{-2}$ and $~10$ g m$^{-2}$ (Fig. \ref{f:probvsr}).  In the stochastic model, the effect of `self-recovery', or escaping the path to extinction, is more appreciable the larger the amount of rainfall. The corresponding probability of extinction for the deterministic system is equal to one in almost the whole regime investigated (not shown). 

Under certain conditions, the opposite effect can happen, with demographic noise promoting extinction in cases in which the deterministic system survives instead. However, this needs a range of {the parameter $\mu$} which we expect not to be relevant for real-world situations. In particular, we have observed such behaviour only for large values of the parameter $\mu$ and for low initial cover. More specifically, we have observed that this can happen if $f\lesssim 10\%$ and $\mu\gtrsim 10$ g m$^{-2}$, for $R \gtrsim 0.5$ mm d$^{-1}$, or $\mu\gtrsim 5$ g m$^{-2}$, for $0.4\lesssim R\lesssim 0.5$ mm d$^{-1}$.

\section{Discussion}

{Two general approaches to modelling ecosystems, and complex systems in general, are continuous models and individual based models. The former tends to be relatively simple and amenable to mathematical analysis, while the latter tends to be more realistic and rely heavily on computational simulations due to its usually higher complexity. In particular, individual based models of ecosystems are mainly stochastic reflecting the} {random behaviour commonly observed in natural systems. On the other hand, continuous models are mostly deterministic or include a noise term added {\it ad hoc}. In some circumstances these can be seen as describing the most relevant features of individual based models. However, it is well known that demographic stochasticity can} have important consequences that are usually neglected when modelling ecosystems in terms of continuous densities  \citep[see e.g.][]{McKane-TREE}. In this work, we represented semi-arid ecosystems with two parallel models: a hybrid stochastic model, including individual-based vegetation dynamics and deterministic water dynamics, and a deterministic model, governing the behaviour of the average stochastic variables. The models represented explicitly the water-vegetation infiltration feedback. Thus, for the parameter values we studied, both models contained solutions corresponding to stable vegetation patterns, analogously to what was observed previously with other models of this type of ecosystem \citep[e.g.][]{Rietkerk-2002}. 

{While continuous models assume that stochastic effects are irrelevant and so neglect them altogether, we have also made} {two assumptions to simplify the computational analysis, i.e. we have assumed the ecosystem is not very sensitive to the growth phase of each individual plant so we can neglect it, and we have taken the same mass for all plants. We expect this to be a reasonable approximation if the time it takes for a plant to establish is much smaller than the time scale on which there is an appreciable change of the spatial distribution of biomass, e.g. the characteristic time for pattern formation.} 

{Conditioned on the validity of these assumptions,} our analysis showed that the resilience of the vegetation patterns changed when taking into account the discrete nature of plants and the intrinsic stochasticity of their behaviour. Including such stochasticity, the modelled ecosystems did not turn into deserts in a wide range of cases in which the deterministic representation would predict the ecosystem to become extinct. This is an important observation, given that semi-arid ecosystems are characterised by a rather scarce number of plants, scattered across regions of empty land. In a finite population, the number of individuals varies because of the intrinsic stochastic behaviour of the individuals. Usually, the fewer the  individuals, the stronger the effects of the demographic noise  \citep{Grima-Nature,Rogers-2012, Biancalani-2011, Goldenfeld-2011, Biancalani-2010,Goldenfeld-2009, McKane-PRL2005}. Intuitively, this is expected to promote extinction when the number of plants is small. It is therefore remarkable that we observed a relevant regime of parameters in which including the stochastic individual nature of the plants actually increased the likelihood of vegetation pattern emergence, and of escaping the desert state. The opposite might also occur, i.e. a larger probability of becoming extinct in the stochastic model, especially when considering large individual plants, covering initially only a small fraction of soil. Under these conditions, the probability of a large stochastic deviation that brings the system suddenly to extinction, could be non-negligible. However, we expect that our approach is better justified for intermediate values of $\mu$, i.e. the biomass of a plant (per unit area), where the results were rather robust. { Stochastic behaviour was not observed when $\mu$ was too small, in which case our model coincided with a deterministic continuous model, whereas if $\mu$ was too large the growth phase and heterogeneity was more noticeable.} When the individual plant biomass was relatively high ($\mu \gtrsim 8$ g m$^{-2}$), or very low ($\mu \to 0$), the outcomes of the stochastic model tended to have higher probability of being desert than for intermediate $\mu$ values. Demographic noise seemed to be more important in an intermediate range of single-plant biomass, which corresponds to a realistic range for herbs or grass biomass \citep[see e.g.][]{Peters-2002}.

In order to study the resilience of the vegetation patterns, we addressed the issue of how and how easily they were reached in time, i.e. for which type of initial conditions the system evolved towards the pattern \citep[e.g.][]{Eppinga2009a}. This gave an indication of how ``attractive'' the patterns were. In the case of the deterministic approximation, this was equivalent to studying the basin of attraction of the patterned stable states. This notion, however, is not applicable to stochastic systems, where there is no notion of equilibrium point, and no deterministic dynamics uniquely leading from a certain initial condition to a final state. For this reason, we focused on the probability of extinction as a function of the initial conditions and for different parameter values. For a large set of initial conditions and parameters, we observed that the stochastic model almost never led the vegetation to extinction, while the vegetation in the deterministic model almost always became extinct. When varying rainfall in a realistic range for drylands, the effect of `self-recovery' in the stochastic model was promoted for the highest rainfall values, where the contrast with the deterministic case was sharper.

{Given the success of deterministic continuous models to date, demographic noise is expected to play a relevant role only when the number of plants is rather low. Indeed we have observed (see e.g. Fig. \ref{f:homo}) that initially the two models follow essentially the same dynamics towards desertification, and only when they are close to extinction the two dynamics diverge: the deterministic dynamics follows the path to extinction, while the stochastic one recovers. This particular} difference between the two models was especially important for vegetation initially covering more than 30-40\%, where the deterministic case would practically always become extinct. {We should be careful, however, not to give too much weight to the importance of the initial conditions, since they were arbitrarily chosen in simulations. The most robust statements of this work had to do with comparing what happened with both the deterministic and stochastic models after they had followed their own dynamics for a while. Nevertheless, such a behaviour of the deterministic model might appear counter-intuitive, in the sense that it might be concluded that {a semi-arid ecosystem with many plants, e.g. due to planting, risks extinction. The stochastic model, on the other hand, did not show this counter-intuitive behaviour since the probability of extinction was {equally small (less than $0.1$) for all values of initial cover. In the case of the deterministic model,} the more homogeneously distributed the plants were, the more difficult it was for any particular vegetation patch to actively increase its own soil water. The water-vegetation feedback, {in this case}, needed contrasts between vegetated and bare patches to take place effectively. For the same reasons, in this regime the homogeneous vegetated state was not stable, and full homogeneous vegetation cover would quickly evolve into a desert state. {Not so for the stochastic model, though, which could escape extinction even in this case}. On the other hand, for very low initial vegetation cover (less than 20-30\%), the results of the deterministic and the stochastic approaches were very similar. We must underline that these low fractions of vegetation cover are quite realistic in the most arid ecosystems.}  One may wonder how it is that for large initial cover (e.g. larger than $40$\%) the deterministic system has a large probability of extinction, while for small cover (e.g. smaller than $20$\%) it almost always survives; after all the dynamics is continuous and to reach extinction the system first needs to go through states of small cover. However, the spatial structure of the intermediate states thus reached does not necessarily coincide with that of a state with homogeneous distribution of soil and surface water and a random distribution of constant biomass $\rho_0$, as were used here as initial conditions. {This stresses the importance of being careful about interpreting our particular choice of initial conditions, as mentioned above.}}}

 The deterministic model we investigated in this work describes exactly the typical behaviour of the stochastic system when the individual plant biomass was negligible ($\mu\to 0$). In this case we could have a very large number of plants per cell. This deterministic model was a good approximation to the kernel model investigated by \cite{Pueyo2008} in the case that the range of the kernel of seed dispersal in the latter was of the order of one or two cells ($\sim 2$-$4$ m). Under these conditions, the kernel could be approximated with an effective diffusive term and with a diffusion coefficient depending on the state of the soil water. If, additionally, we could neglect the spatial variation of the soil water, the two models became similar to the model studied by \cite{Rietkerk-2001}. In this sense, we could say that the stochastic model introduced here was close to these well-studied models and, in particular, it was not unexpected to observe spatial patterns in a similar range of parameters. What could change drastically is the transitory dynamics while reaching a stationary state, as we showed with the effect of self-recovery. 

In principle, a more complete approach would have to deal with each single plant individually, with all its attributes and ongoing processes, as has been done for instance in the study of forests, see e.g. \cite{Perry-2006}. However, this would make the problem far more complex from a computational point of view, and it might become difficult to gain insight. Indeed, as has been discovered in the investigations on forests, a far simpler analytical approximation, called the perfect plastic approximation, may capture most of the relevant details of the dynamics observed in simulations \citep{Pacala-2008}. This case, however, corresponds to a regime of high vegetation density, where fluctuations of the average behaviour are expected to be irrelevant.  This raises the question of which would be the `best' way of modelling a plant in the study of semi-arid ecosystems in the sense of, paraphrasing Einstein, `keeping it simple, but not too simple'. 

In our stochastic modelling approach, we did not consider the heterogeneity in plant size. In particular, we did not include the different plant life stages. In a sense, plants were instantaneously created as adult individuals. This might appear as an unrealistic feature that might promote the effect of `self-recovery' we discussed, because a plant could start increasing the availability of water locally as soon as it was created, and therefore instantly start promoting its own survival. However, this feature would also lead to an overestimation of water uptake. Since in our model plants died suddenly, in detriment of self-recovery, mortality was also over-estimated. Furthermore, in the stochastic model we investigated, a plant could produce another plant only in the neighbouring cells, thus limiting the impact the new birth had on the state of the ecosystem. 

A next step in the complexity of the modelling could be to introduce two type of individuals: seedlings and established plants.  In this way we would be able to take into account, for instance, the high asymmetry in mortality between these two. The question is then how robust are the results we have discussed in this work under this more realistic scenario. At first sight, one could expect that the stochastic model becomes closer to the deterministic approximation, but experience in this field of research has shown that counter-intuitive effects are not uncommon \citep{Grima-Nature,Rogers-2012, Biancalani-2011, Goldenfeld-2011, Biancalani-2010,Goldenfeld-2009, McKane-PRL2005}.  

The model we presented did not include environmental heterogeneity and stochasticity, or topography \citep[see e.g.][]{Sheffer-2013}. We also discarded rainfall intermittence. Vegetation in arid areas is well adapted to the occasional occurrence of rainfall \citep[][]{Baudena2007,Kletter2009, Dodorico2007}, although the effect of rainfall intermittency may not be too relevant when spatial feedbacks are represented \citep[see][]{Baudena2008a}. Besides water, we did not consider any other limiting resource, such as nutrient or light limitations. Moreover, we did not include another water-vegetation feedback mechanism, which is known to play a role in drylands, namely the effect of root length. Plant water availability increases with the root extent, which in turn increases with the biomass itself, thus favouring plant growth and generating vegetation patterns \citep{Gilad2004,Lefever1997,Barbier2008}.

Another relevant issue is how to validate our findings with observations. For example, we could analyse patch dynamics to see whether it displays the `wiggly' behaviour observed in the model results. This nevertheless would require time series of spatial patterns, with an appropriate spatial and temporal resolution, which may not be easily obtainable. In principle, it could even be possible to compute the statistics of this patch dynamics, in order to have quantitative predictions.

Despite not being conclusive in any sense, this investigation indicated that, in certain regimes, including demographic noise could lead to a larger estimate of the resilience of semi-arid ecosystems. Our model results suggested that demographic noise may be more important in the less arid ecosystems, with higher rainfall and vegetation cover. Since changes in rainfall regimes are expected, for example as a consequence of climate change, it may be necessary to take into account  individual-based dynamics to evaluate the resilience and resistance of these ecosystems to such forcing. In summary, we think the study of semi-arid ecosystems might benefit from the approach taken for instance in the research on forests, where quite detailed IBM's have been extensively used. Indeed, in contrast to forests which are characterised by a rather dense vegetation, the typical number of plants in semi-arid ecosystems is comparatively quite low and so the stochastic effects implicit in such a modelling approach are expected to be more relevant.

\section*{Acknowledgements}
The authors thank M. Eppinga for useful discussions. This research is supported by the ERA Complexity Net through the RESINEE project (`Resilience and interaction of networks in ecology and economics'). UK support for this project to JRG, TG and AJM is administered by the Engineering and Physical Sciences Research Council EPSRC (grant reference EP/I019200/1). TG acknowledges funding by RCUK (EP/E500048/1). The research of MB and MR is also supported by the project CASCADE (Seventh Framework Programme FP7/2007-2013 grant agreement 283068).

\clearpage

\begin{thebibliography}{60}
\expandafter\ifx\csname natexlab\endcsname\relax\def\natexlab#1{#1}\fi
\expandafter\ifx\csname url\endcsname\relax
  \def\url#1{\texttt{#1}}\fi
\expandafter\ifx\csname urlprefix\endcsname\relax\def\urlprefix{URL }\fi

\bibitem[{Aguiar and Sala(1999)}]{Aguiar1999}
Aguiar, M., Sala, O.~E., 1999. Patch structure, dynamics and implications for
  the functioning of arid ecosystems. Trends in Ecology \& Evolution 14,
  273--277.

\bibitem[{Barbier et~al.(2008)Barbier, Couteron, Lefever, Deblauwe, and
  Lejeune}]{Barbier2008}
Barbier, N., Couteron, P., Lefever, R., Deblauwe, V., Lejeune, O., 2008.
  Spatial decoupling of facilitation and competition at the origin of gapped
  vegetation patterns. Ecology 89, 1521--31.

\bibitem[{Barbier et~al.(2006)Barbier, Couteron, Lejoly, Deblauwe, and
  Lejeune}]{Barbier2006}
Barbier, N., Couteron, P., Lejoly, J., Deblauwe, V., Lejeune, O., 2006.
  Self-organized vegetation patterning as a fingerprint of climate and human
  impact on semi-arid ecosystems. Journal of Ecology 94, 537--547.

\bibitem[{Baudena et~al.(2007)Baudena, Boni, Ferraris, von Hardenberg, and
  Provenzale}]{Baudena2007}
Baudena, M., Boni, G., Ferraris, L., von Hardenberg, J., Provenzale, A., 2007.
  Vegetation response to rainfall intermittency in drylands: {R}esults from a
  simple ecohydrological box model. Advances in Water Resources 30, 1320--1328.

\bibitem[{Baudena and Provenzale(2008)}]{Baudena2008a}
Baudena, M., Provenzale, A., 2008. Rainfall intermittency and vegetation
  feedbacks in drylands. Hydrology and Earth System Sciences 12, 679--689.

\bibitem[{Baudena and Rietkerk(2012)}]{Baudena2012b}
Baudena, M., Rietkerk, M., 2012. Complexity and coexistence in a simple spatial
  model for arid savanna ecosystems. Theoretical Ecology.

\bibitem[{Biancalani et~al.(2010)Biancalani, Fanelli, and
  Patti}]{Biancalani-2010}
Biancalani, T., Fanelli, D., Patti, F.~D., 2010. Stochastic {T}uring patterns
  in a {B}russelator model. Physical Review E 81, 046215.

\bibitem[{Biancalani et~al.(2011)Biancalani, Galla, and
  McKane}]{Biancalani-2011}
Biancalani, T., Galla, T., McKane, A.~J., 2011. Stochastic waves in a
  {B}russelator model with nonlocal interactions. Physical Review E 84, 026201.

\bibitem[{Black and McKane(2011)}]{McKane-WKB-2011}
Black, A.~J., McKane, A.~J., 2011. {WKB} calculation of an epidemic outbreak
  distribution. Journal of Statistical Mechanics, P12006.

\bibitem[{Black and McKane(2012)}]{McKane-TREE}
Black, A.~J., McKane, A.~J., 2012. Stochastic formulation of ecological models
  and their applications. Trends in Ecology \& Evolution 27, 337--345.

\bibitem[{Botkin et~al.(1972)Botkin, Janak, and Wallis}]{Botkin-1972}
Botkin, D.~B., Janak, J.~F., Wallis, J.~R., 1972. Rationale, limitations, and
  assumptions of a northeastern forest growth simulator. IBM Journal of
  Research \& Development 16, 101--116.

\bibitem[{Butler and Goldenfeld(2009)}]{Goldenfeld-2009}
Butler, T., Goldenfeld, N., 2009. Robust ecological pattern formation induced
  by demographic noise. Physical Review E 80, 030902(R).

\bibitem[{Butler and Goldenfeld(2011)}]{Goldenfeld-2011}
Butler, T., Goldenfeld, N., 2011. Fluctuation-driven {T}uring patterns.
  Physical Review E 84, 011112.

\bibitem[{Casenave and Valentin(1992)}]{Casenave-1992}
Casenave, A., Valentin, C., 1992. A runoff capability classification system based on surface features criteria in semi-arid areas of West Africa. Journal of Hydrology 130, 231--249.

\bibitem[{Coffin and Lauenroth(1990)}]{Coffin-1990}
Coffin, D., Lauenroth, W., 1990. A gap dynamics simulation model of succession
  in a semiarid grassland. Ecological Modelling 49, 229--266.

\bibitem[{Crawley(1990)}]{Crawley-1990}
Crawley, M.~J., 1990. The population dynamics of plants. Phil. Trans. R. Soc.
  Lond. B 330, 125--140.

\bibitem[{Cross and Greenside(2009)}]{Cross-book}
Cross, M.~C., Greenside, H.~S., 2009. Pattern formation and dynamics in
  non-equilibrium systems. Cambridge University Press, Cambridge.

\bibitem[{Dakos et~al.(2011)Dakos, K\'efi, Rietkerk, van Nes, and
  Scheffer}]{Rietkerk-2012}
Dakos, V., K\'efi, S., Rietkerk, M., van Nes, E.~H., Scheffer, M., 2011.
  Slowing down in spatially patterned ecosystems at the brink of collapse. The
  American Naturalist 177, E155--E166.

\bibitem[{Davis(1984)}]{Davis-1984}
Davis, M. H.~A., 1984. Piecewise-deterministic {M}arkov processes: a general
  class of non-diffusion stochastic models (with discussion). J. R. Stat. Soc.
  B 46, 353--388.

\bibitem[{Deblauwe et~al.(2008)Deblauwe, Barbier, Couteron, Lejeune, and
  Bogaert}]{Deblauwe2008}
Deblauwe, V., Barbier, N., Couteron, P., Lejeune, O., Bogaert, J., 2008. The
  global biogeography of semi-arid periodic vegetation patterns. Global Ecology
  and Biogeography 17, 715--723.

\bibitem[{D'Odorico et~al.(2007)D'Odorico, Laio, Porporato, Ridolfi, and
  Barbier}]{Dodorico2007}
D'Odorico, P., Laio, F., Porporato, A., Ridolfi, L., Barbier, N., 2007.
  Noise-induced vegetation patterns in fire-prone savannas. Journal of
  Geophysical Research 112, G02021--G02021.

\bibitem[{Eppinga(2009)}]{Eppinga2009a}
Eppinga, M., 2009. {Amazing pattern}. Ph.D. thesis, Universiteit Utrecht.

\bibitem[{Faggionato et~al.(2009)Faggionato, Gabrielli, and
  Crivellari}]{Faggionato-2009}
Faggionato, A., Gabrielli, D., Crivellari, M.~R., 2009. Non-equilibrium
  thermodynamics of piecewise deterministic {M}arkov processes. Journal of
  Statistical Physics 137, 259--304.

\bibitem[{Faggionato et~al.(2010)Faggionato, Gabrielli, and
  Crivellari}]{Faggionato-2010}
Faggionato, A., Gabrielli, D., Crivellari, M.~R., 2010. Averaging and large
  deviation principles for fully-coupled piecewise deteministic {M}arkov
  processes and applications to molecular motors. Markov Processes Related
  Fields 16, 497--548.

\bibitem[{Frey(2010)}]{Frey-2010}
Frey, E., 2010. Evolutionary game theory: Theoretical concepts and applications
  to microbial communities. Physica A: Statistical Mechanics and its
  Applications 389, 4265--4298, section 2.1.

\bibitem[{Gilad et~al.(2004)Gilad, von Hardenberg, Provenzale, Shachak, and
  Meron}]{Gilad2004}
Gilad, E., von Hardenberg, J., Provenzale, A., Shachak, M., Meron, E., 2004.
  Ecosystem engineers: from pattern formation to habitat creation. Physical
  Review Letters 93, 98105.

\bibitem[{Gillespie(1976)}]{Gillespie-1976}
Gillespie, D.~T., 1976. A general method for numerically simulating the
  stochastic time evolution of coupled chemical reactions. Journal of
  Computational Physics 22, 403--434.

\bibitem[{Gillespie(1977)}]{Gillespie-1977}
Gillespie, D.~T., 1977. Exact stochastic simulation of coupled chemical
  reactions. Journal of Physical Chemistry 81, 2340--2361.

\bibitem[{Grimm and Calabrese(2011)}]{Grimm-2011}
Grimm, V., Calabrese, J.~M., 2011. What is resilience? {A} short introduction.
  In: Deffuant, G., Gilbert, N. (Eds.), Viability and resilience of complex
  systems: concepts, methods and case studies from ecology and society.
  Springer, Heidelberg, Ch.~1, p.~15.

\bibitem[{HilleRisLambers et~al.(2001)HilleRisLambers, Rietkerk, van~den Bosch,
  Prins, and de~Kroon}]{Rietkerk-2001}
HilleRisLambers, R., Rietkerk, M., van~den Bosch, F., Prins, H. H.~T.,
  de~Kroon, H., 2001. Vegetation pattern formation in semi-arid grazing
  systems. Ecology 82, 50–--61.

\bibitem[{Holling(1973)}]{Holling-1973}
Holling, C., 1973. Resilience and stability of ecological systems. Annual
  Review of Ecology and Systematics 4, 1--23.

\bibitem[{J{\o}rgensen(2011)}]{Jorgensen-Handbook}
J{\o}rgensen, S.~E. (Ed.), 2011. Handbook of Ecological Models used in
  Ecosystem and Environmental Management. {CRC} {P}ress, Boca Raton.

\bibitem[{K\'{e}fi et~al.(2007{\natexlab{a}})K\'{e}fi, Rietkerk, Alados, Pueyo,
  Papanastasis, Elaich, and de~Ruiter}]{Kefi2007}
K\'{e}fi, S., Rietkerk, M., Alados, C.~L., Pueyo, Y., Papanastasis, V.~P.,
  Elaich, A., de~Ruiter, P.~C., 2007{\natexlab{a}}. Spatial vegetation patterns
  and imminent desertification in {M}editerranean arid ecosystems. Nature 449,
  213--217.

\bibitem[{K\'{e}fi et~al.(2007{\natexlab{b}})K\'{e}fi, Rietkerk, van Baalen,
  and Loreau}]{Kefi2007a}
K\'{e}fi, S., Rietkerk, M., van Baalen, M., Loreau, M., 2007{\natexlab{b}}.
  Local facilitation, bistability and transitions in arid ecosystems.
  Theoretical {P}opulation {B}iology 71, 367--379.

\bibitem[{Klausmeier(1999)}]{Klausmeier1999}
Klausmeier, C., 1999. Regular and irregular patterns in semi-arid vegetation.
  Science 284, 1826--1828.

\bibitem[{Kletter et~al.(2009)Kletter, von Hardenberg, Meron, and
  Provenzale}]{Kletter2009}
Kletter, A.~Y., von Hardenberg, J., Meron, E., Provenzale, A., 2009. Patterned
  vegetation and rainfall intermittency. Journal of Theoretical Biology 256,
  574--583.

\bibitem[{{Lefever, R. and Lejeune}(1997)}]{Lefever1997}
{Lefever, R. and Lejeune}, O., 1997. On the origin of tiger bush. Bulletin of
  Mathematical Biology 59, 263--294.

\bibitem[{Manor and Shnerb(2008)}]{Manor2008}
Manor, A., Shnerb, N.~M., 2008. Facilitation, competition, and vegetation
  patchiness: from scale free distribution to patterns. Journal of Theoretical
  Biology 253, 838--842.

\bibitem[{Martin and Calabrese(2011)}]{Martin-2011}
Martin, S., Calabrese, G. D. J.~M., 2011. Defining resilience mathematically:
  from attractors to viability. In: Deffuant, G., Gilbert, N. (Eds.), Viability
  and resilience of complex systems: concepts, methods and case studies from
  ecology and society. Springer, Heidelberg, Ch.~2, p.~3.

\bibitem[{McKane and Newman(2005)}]{McKane-PRL2005}
McKane, A.~J., Newman, T.~J., 2005. Predator-prey cycles from resonant
  amplification of demographic stochasticity. Physical Review Letters 94,
  218102.

\bibitem[{Meron(2012)}]{Meron2012}
Meron, E., 2012. Pattern-formation approach to modelling spatially extended
  ecosystems. Ecological Modelling 234, 70--82.

\bibitem[{Meron(2011)}]{Meron-2011}
Meron, E., 2011. Modeling dryland landscapes. Mathematical Modelling of Natural Phenomena 6, 163--187.

\bibitem[{Moorcroft et~al.(2001)Moorcroft, Hurtt, and Pacala}]{Moorcroft2001}
Moorcroft, P.~R., Hurtt, G.~C., Pacala, S.~W., 2001. A method for scaling
  vegetation dynamics: the ecosystem demography model ({ED}). Ecological
  Monographs 71, 557--586.

\bibitem[{Pacala et~al.(1996)Pacala, Canham, Saponara, Jr., Kobe, and
  Ribbens}]{Pacala-1996}
Pacala, S.~W., Canham, C.~D., Saponara, J., Jr., J. A.~S., Kobe, R.~K.,
  Ribbens, E., 1996. Forest models defined by field measurements: Estimation,
  error analysis and dynamics. Ecological Monographs 66, 1--43.

\bibitem[{Pacala et~al.(1993)Pacala, Canham, and Silander~Jr.}]{Pacala-1993}
Pacala, S.~W., Canham, C.~D., Silander~Jr., J.~A., 1993. Forest models defined
  by field measurements: I. the design of a northeastern forest simulator.
  Canadian Journal of Forest Research 23, 1980--1988.

\bibitem[{Perry and Enright(2006)}]{Perry-2006}
Perry, G. L.~W., Enright, N.~J., 2006. Spatial modelling of vegetation change
  in dynamic landscapes: a review of methods and applications. Progress in
  Physical Geography 30, 47--72.

\bibitem[{Peters(2002)}]{Peters-2002}
Peters, D. P.~C., 2002. Plant species dominance at a grassland-shrubland
  ecotone: an individual-based gap dynamics model of herbaceous and woody
  species. Ecological Modelling 152, 5--32.

\bibitem[{Pueyo et~al.(2008)Pueyo, K\'efi, Alados, and Rietkerk}]{Pueyo2008}
Pueyo, Y., K\'efi, S., Alados, C.~L., Rietkerk, M., 2008. Dispersal strategies
  and spatial organization of vegetation in arid ecosystems. Oikos 117,
  1522--1532.

\bibitem[{Ramaswamy et~al.(2012)Ramaswamy, Gonzalez-Segredo, Sbalzarini, and
  Grima}]{Grima-Nature}
Ramaswamy, R., Gonzalez-Segredo, N., Sbalzarini, I.~F., Grima, R., 2012.
  Discreteness-induced concentration inversion in mesoscopic chemical systems.
  Nature Communications 3, 779.

\bibitem[{Rastetter et~al.(2003)Rastetter, Aber, Peters, Ojima, and
  Burke}]{Rastetter-2003}
Rastetter, E.~B., Aber, J.~D., Peters, D. P.~C., Ojima, D.~S., Burke, I.~C.,
  2003. Using mechanistic models to scale ecological processes across space and
  time. Bio{S}cience 53, 68--76.

\bibitem[{Realpe-Gomez et~al.(2012)Realpe-Gomez, Galla, and
  McKane}]{McKane-2012}
Realpe-Gomez, J., Galla, T., McKane, A.~J., 2012. Demographic noise and
  piecewise deterministic {M}arkov processes. Physical Review E 86, 011137.

\bibitem[{Rietkerk et~al.(2002)Rietkerk, Boerlijst, van Langevelde,
  Hille{R}is{L}ambers, van~de Koppel, Kumar, Prins, and
  de~Roos}]{Rietkerk-2002}
Rietkerk, M., Boerlijst, M.~C., van Langevelde, F., Hille{R}is{L}ambers, R.,
  van~de Koppel, J., Kumar, L., Prins, H. H.~T., de~Roos, A.~M., 2002.
  Self-organization of vegetation in arid ecosystems. American Naturalist 160,
  524--530.

\bibitem[{Rietkerk et~al.(2004)Rietkerk, Dekker, de~Ruiter, and van~de
  Koppel}]{Rietkerk2004}
Rietkerk, M., Dekker, S.~C., de~Ruiter, P.~C., van~de Koppel, J., 2004.
  Self-organized patchiness and catastrophic shifts in ecosystems. Science 305,
  1926--1929.

\bibitem[{Rietkerk and van~de Koppel(1997)}]{Rietkerk1997}
Rietkerk, M., van~de Koppel, J., 1997. Alternate stable states and threshold
  effects in semi-arid grazing systems. Oikos 79, 69--76.

\bibitem[{Rietkerk and van~de Koppel(2008)}]{Rietkerk2008}
Rietkerk, M., van~de Koppel, J., 2008. Regular pattern formation in real
  ecosystems. Trends in Ecology \& Evolution 23, 169--175.

\bibitem[{Rogers et~al.(2012)Rogers, McKane, and Rossberg}]{Rogers-2012}
Rogers, T., McKane, A.~J., Rossberg, A.~G., 2012. Demographic noise can lead to
  the spontaneous formation of species. Europhysics Letters 97, 40008.

\bibitem[{San Miguel et~al.(2012)San Miguel, Johnson, Kertesz, Kaski, Díaz-Guilera, MacKay, Loreto, Érdi, and Helbing}]{SanMiguel-2012}
San Miguel, M.,	Johnson, J. H., Kertesz, J., Kaski, K., Díaz-Guilera, A., MacKay, R.S., Loreto, V., Érdi, P., Helbing, D., 2012. Challenges in complex systems science. 
  The European Physical Journal Special Topics 214, 245--271.

\bibitem[{Schenk et~al.(2008)Schenk, Espino, Goedhart, Nordenstahl, Martinez Cabrera, and Jones}]{Schenk-2008}
Schenk, H. J., Espino, S., Goedhart, C. M., Nordenstahl, M., Martinez Cabrera, H. I., Jones, C. S., 2008. Hydraulic integration and shrub growth form linked across continental aridity gradients. 
  Proceedings of the National Academy of Sciences of The United States of America 105, 11248--11253.

\bibitem[{Sheffer et~al.(2013)Sheffer, von Hardenberg, Yizhaq, Shachak, and Meron}]{Sheffer-2013}
Sheffer, E.,  von Hardenberg, J., Yizhaq, H., Shachak, M., Meron, E., 2013. Emerged or imposed: a theory on the role of physical templates and self-organisation for vegetation patchiness.
  Ecology Letters 16, 127--139.

\bibitem[{Shugart(1984)}]{Shugart-1984}
Shugart, H., 1984. A theory of forest dynamics. Springer-Verlag, New York.

\bibitem[{van der Stelt et~al.(2013)van der Stelt, Doelman, Hek, and
  Rademacher}]{Stelt-2013}
van der Stelt, S., Doelman, A., Hek, G., Rademacher, J. D. M., 2013. Rise and fall of periodic patterns for a generalized Klausmeier-Gray-Scott model.
  Journal of Nonlinear Science 23, 39--95.

\bibitem[{Strigul et~al.(2008)Strigul, Pristinski, Purves, Dushoff, and
  Pacala}]{Pacala-2008}
Strigul, N., Pristinski, D., Purves, D., Dushoff, J., Pacala, S., 2008. Scaling
  from trees to forests: tractable macroscopic equations for forest dynamics.
  Ecological Monographs 78, 523--545.

\bibitem[{Turing(1952)}]{Turing-1952}
Turing, A.~M., 1952. The chemical basis of morphogenesis. Phil. Trans. R. Soc.
  B (London) 237, 37--72.

\bibitem[{Vincenot et~al.(2010)Vincenot, Giannino, Rietkerk, Moriya, and
  Mazzoleni}]{Vincenot2010}
Vincenot, C.~E., Giannino, F., Rietkerk, M., Moriya, K., Mazzoleni, S., 2010.
  Theoretical considerations on the combined use of system dynamics and
  individual-based modeling in ecology. Ecological Modelling 222, 210--218.

\bibitem[{von Hardenberg et~al.(2001)von Hardenberg, Meron, Shachak, and
  Zarmi}]{vonHardenberg2001}
von Hardenberg, J., Meron, E., Shachak, M., Zarmi, Y., 2001. Diversity of
  vegetation patterns and desertification. Physical Review Letters 87, 198101.

\bibitem[{Walker et~al.(1981)Walker, Ludwig, Holling, and
  Peterman}]{Walker1981}
Walker, B., Ludwig, D., Holling, C., Peterman, R., 1981. Stability of semi-arid
  savanna grazing systems. Journal of Ecology 69, 473--498.

\end{thebibliography}

\clearpage

\clearpage

\bfi
\includegraphics[width=8cm]{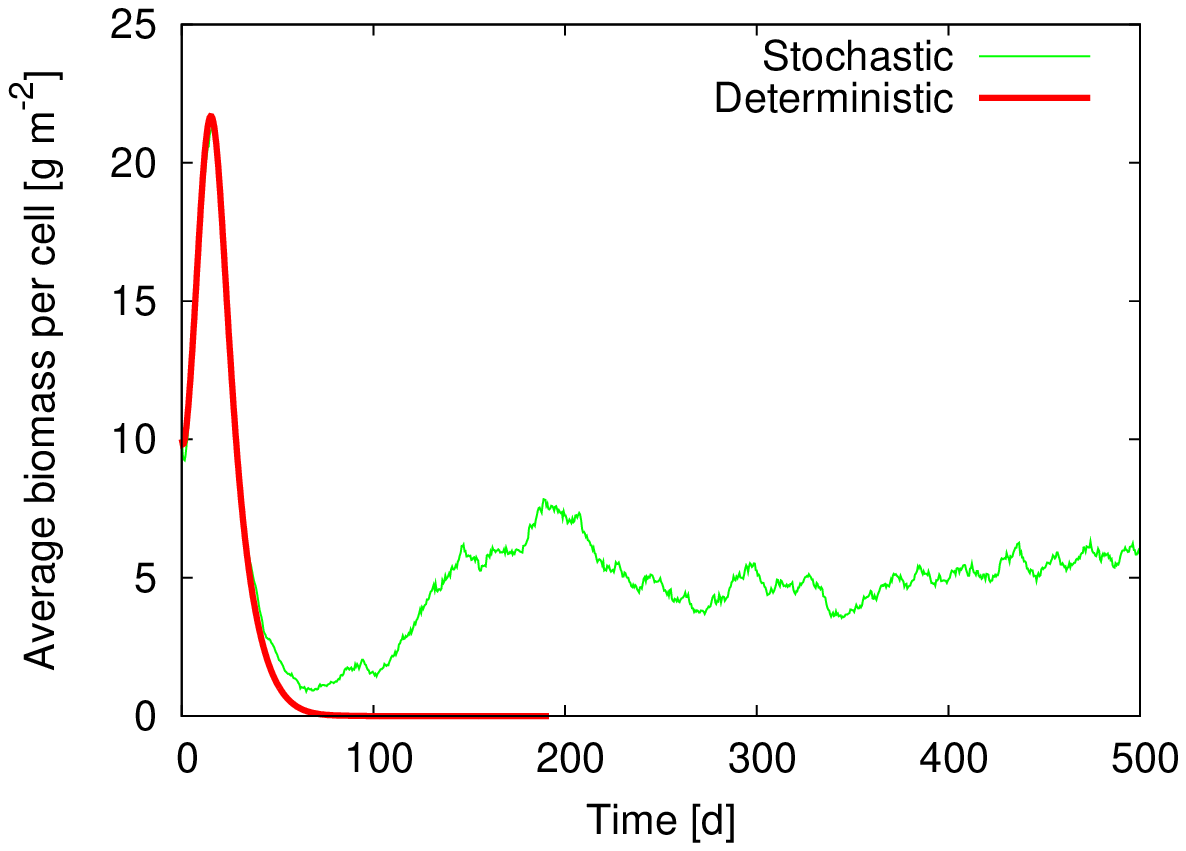}
\caption{(Colour online) Comparison of the dynamics of the total biomass density in the stochastic model (continuous green line) with that in the corresponding deterministic approximation (thick red line). While the deterministic approximation leads to extinction, the full stochastic model recovers. Simulations were run on a line with $128$ cells, periodic boundary conditions and the same random initial conditions in all cells: soil and surface water depths were given by the values in the desert state, Eq. (\ref{e:dfp}), while the initial biomass was $\rho_0 =10$ g m$^{-2}$ ($f=1$). Other key parameter values were: $\mu=1$ g m$^{-2}$, $R=0.6$ mm d$^{-1}$. See Table \ref{t} for the remaining parameter values.}\label{f:homo}
\efi
	
\bfi
\includegraphics[width=9.0cm]{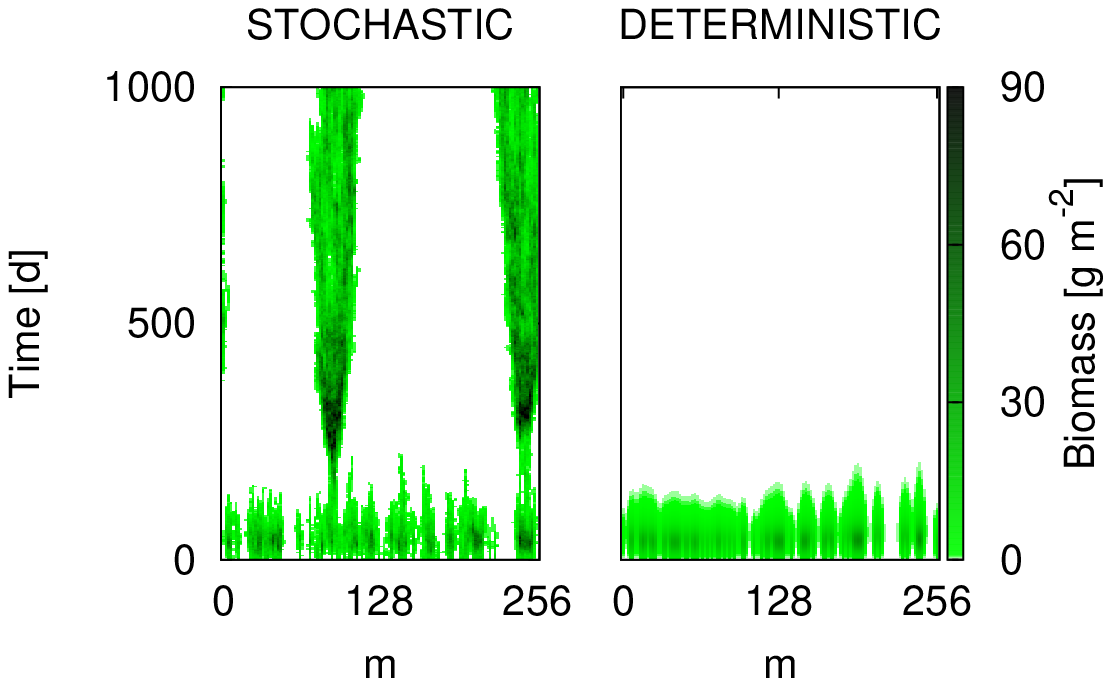}
\caption{(Colour online) Comparison of the dynamics of the vegetation profile in the stochastic model (left) with that in the corresponding deterministic approximation (right). While the deterministic approximation leads to extinction, the full stochastic model recovers. Simulations were run on a line with $128$ cells, periodic boundary conditions and the same initial conditions for both the stochastic model and its deterministic approximation: soil and surface water depths in all cells were given by the values in the desert state, Eq. (\ref{e:dfp}), while biomass was $\rho_0 =0$ g m$^{-2}$ in half of the cells, chosen randomly, and $\rho_0=10$ g m$^{-2}$ in the other half ($f=1/2$). Key parameter values were: $\mu=1$ g m$^{-2}$, $R=0.6$ mm d$^{-1}$. See Table \ref{t} for the remaining parameter values. }\label{f:patternzoom}
\efi

\bfi
\includegraphics[width=10cm, trim= 0mm 0mm 0mm 0mm]{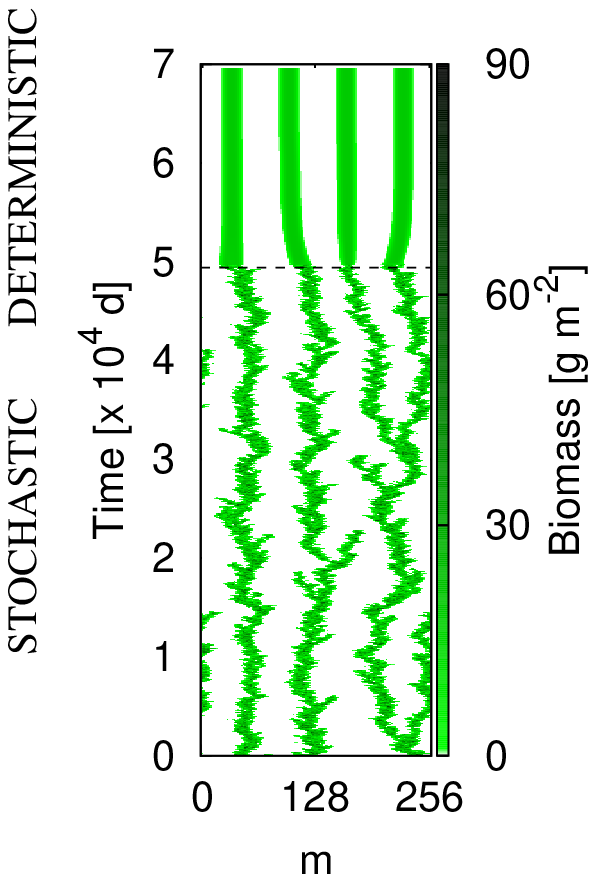}
\caption{(Colour online) Dynamics of the vegetation profile presented on the left side of Fig. \ref{f:patternzoom} (stochastic model) for a longer period of time. Below the horizontal dashed line, i.e. from $t=0$ days to $t=5\times 10^4$ days, we show the dynamical behaviour of the stochastic model, and observe that the system indeed escapes the path to extinction and finally reaches a stationary pattern. Vegetation patches, however, appear to follow dynamics on their own: they can split, merge, and become extinct. In order to compare this with the dynamics of the deterministic model, we use the configuration reached by the stochastic dynamics at $t=5\times 10^4$ days (horizontal dashed line) as an initial condition for the deterministic model. The outcome of the corresponding deterministic dynamics are shown above the horizontal dashed line, i.e. from $t=5\times 10^4$ days to $t=7\times 10^4$ days. In other words, at time $t=5\times 10^4$ days we switch the dynamics from the stochastic to the deterministic model. Clearly the patterns remain stable under the deterministic dynamics. Simulations were run on a line with $128$ cells, periodic boundary conditions and the same initial conditions for both the stochastic model and its deterministic approximation: soil and surface water depths in all cells were given by the values in the desert state, Eq. (\ref{e:dfp}), while biomass was $\rho_0 =0$ g m$^{-2}$ in half of the cells, chosen randomly, and $\rho_0=10$ g m$^{-2}$ in the other half ($f=1/2$). Key parameter values were: $\mu=1$ g m$^{-2}$, $R=0.6$ mm d$^{-1}$. See Table \ref{t} for the remaining parameter values.}\label{f:pattern}
\efi

\bfi
\centering
\vspace{-3cm}
\includegraphics[width=8.5cm]{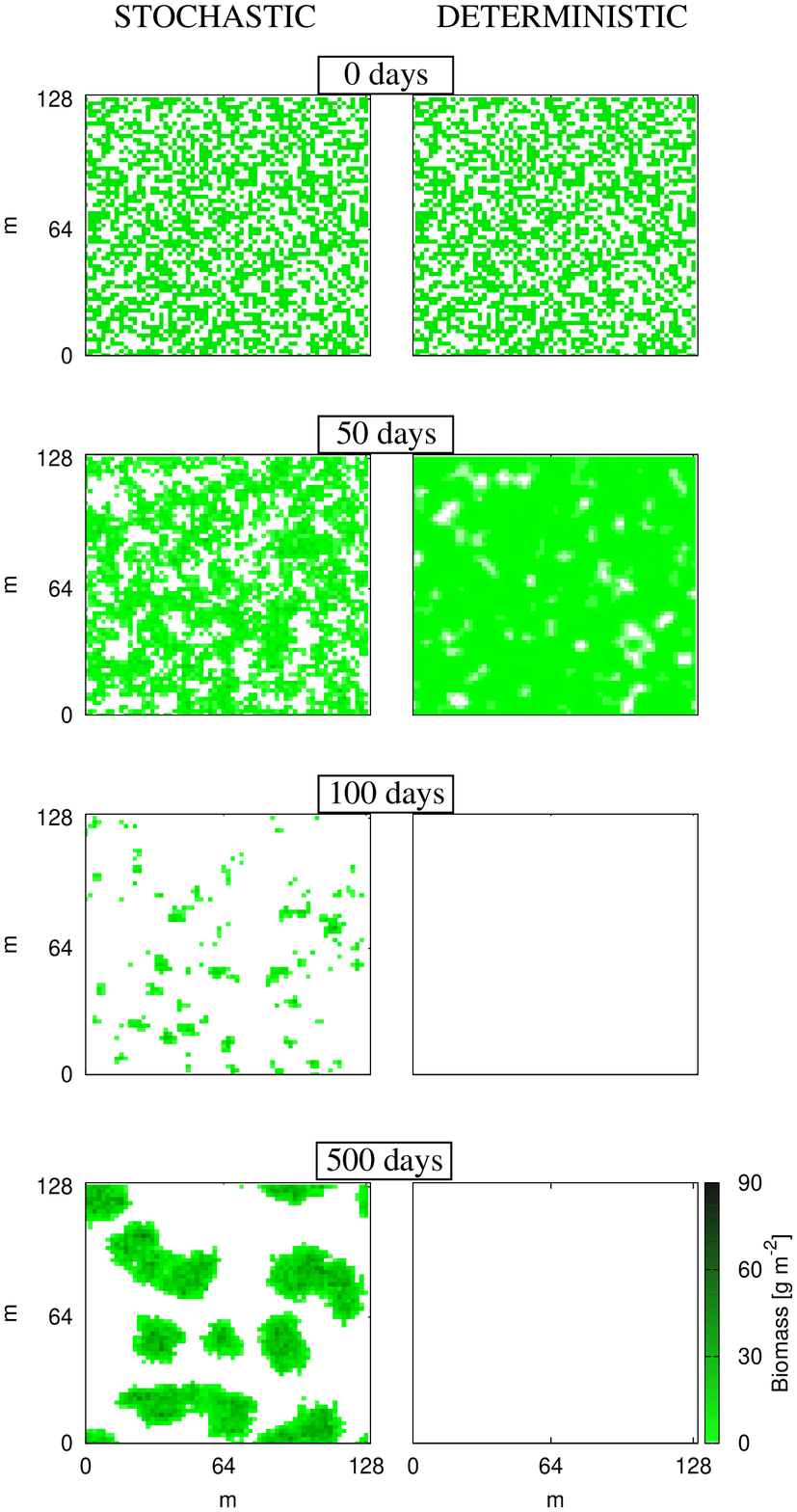}
\caption{(Colour online)  Comparison of the dynamics of the vegetation profile in the stochastic model in two dimensions (left) with that in the corresponding deterministic approximation (right); the axes correspond to the two spatial dimensions. See also the supplementary video which shows the full dynamics of these same vegetation profiles. While the deterministic approximation leads to extinction, the full stochastic model recovers. The simulations were run on a square grid of $64\times 64$ cells and with periodic boundary conditions. Both the stochastic model and its deterministic approximation were started with  the same initial conditions: soil and surface water depths in all cells were given by the values in the desert state, Eq. (\ref{e:dfp}), while biomass was $\rho_0 =0$ g m$^{-2}$ in half of the cells, chosen randomly, and $\rho_0=10$ g m$^{-2}$ in the other half ($f=1/2$). Key parameter values were: $\mu=1$ g m$^{-2}$, $R=0.6$ mm d$^{-1}$. See Table \ref{t} for the remaining parameter values.}\label{f:pattern2d}
\efi

\bfi
\includegraphics[width=8cm]{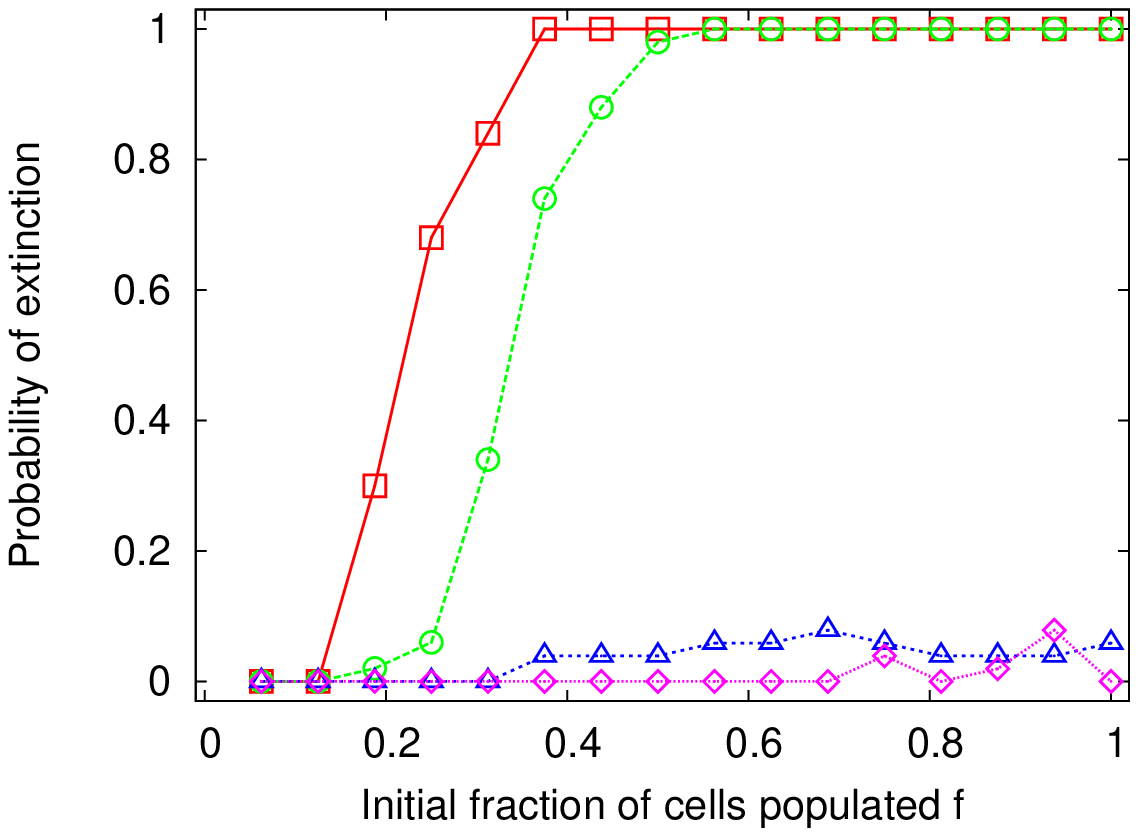}
\caption{(Colour online) Probability of extinction in the stochastic model (blue triangles and magenta rhombi) and in the corresponding approximation to a deterministic model (red squares and green circles). We can observe a very strong contrast between the two: while the deterministic model almost always becomes extinct for cover values $f\gtrsim 0.4$, the stochastic system almost never does, for any value of $f$.  For the stochastic case, the probabilities were estimated from $50$ numerical simulations for each point. The horizontal axis indicates the initial fraction $f$ of populated cells, i.e. cells with initial biomass $\rho_0 >0$. Simulations were run on a line with $128$ cells and with periodic boundary conditions. Both the stochastic model and its deterministic approximation were started with the same initial conditions: in all cells soil and surface water depths were given by the values in the desert state, Eq. (\ref{e:dfp}), while biomass was $\rho_0 =10$ g m$^{-2}$ (green circles and magenta rhombi), and $\rho_0= 50$ g m$^{-2}$ (red squares and blue triangles) in a fraction $f$ of randomly chosen cells and zero in the remaining cells. Key parameter values were: $\mu=1$ g m$^{-2}$, $R=0.6$ mm d$^{-1}$. See Table \ref{t} for the remaining parameter values. }\label{f:probvsf}
\efi

\bfi
\includegraphics[width=8cm]{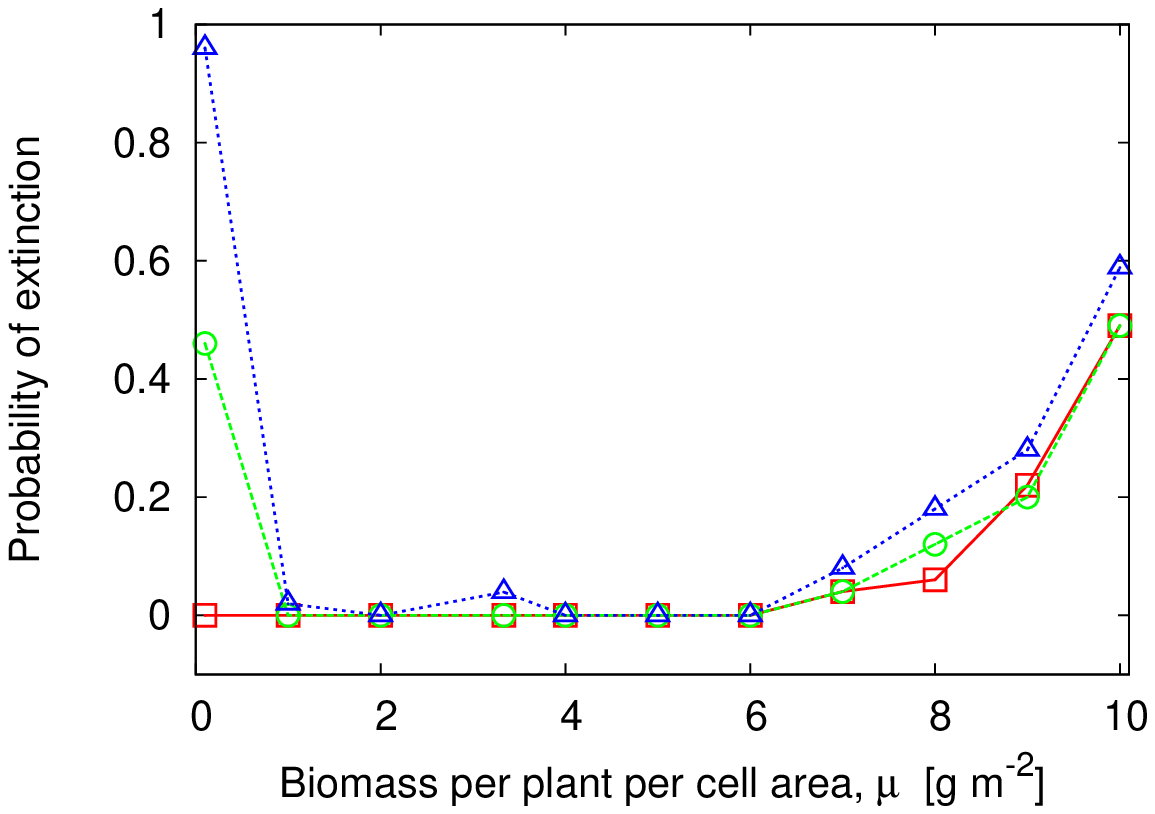}
\caption{(Colour online)   Probability of extinction in the stochastic model as a function of the parameter $\mu$, i.e. the average mass of a plant divided by the area of a cell. The probabilities were estimated from $50$ numerical simulations for each point. Simulations were run on a line with $128$ cells with periodic boundary conditions and initial conditions chosen as follows: in all cells initial soil and surface water depths were given by the values in the desert state, Eq. (\ref{e:dfp}), while biomass was $\rho_0\approx 10$ g m$^{-2}$ in a fraction $f$ of randomly chosen cells and zero in the remaining cells. The three curves correspond to three different values of $f$: $1/8$ (red squares), $1/2$ (green circles) and $7/8$ (blue triangles). See Table \ref{t} for the remaining parameter values. For the same parameter values and initial conditions the probability of extinction in the deterministic model is essentially one throughout the whole regime investigated (not shown). }\label{f:probvsmp}
\efi

\bfi
\includegraphics[width=8cm]{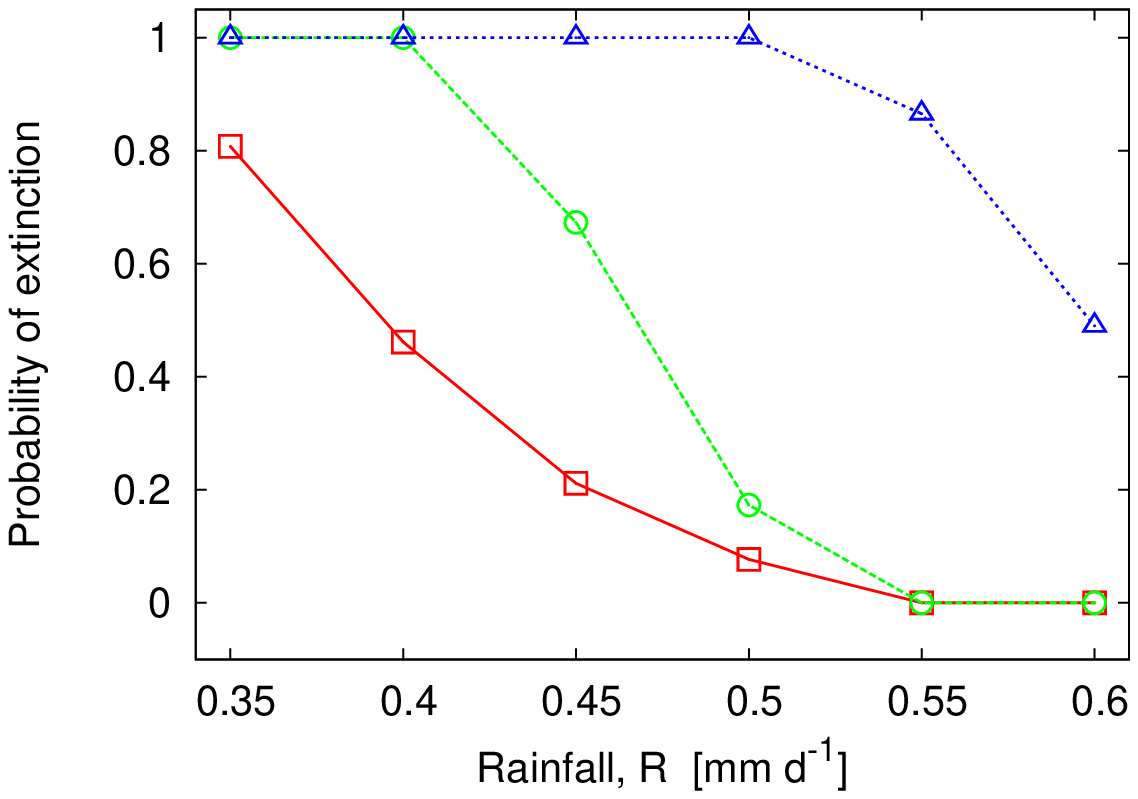}
\caption{(Colour online)  Probability of extinction in the stochastic model as a function of the rainfall rate $R$. Extinction probabilities were estimated from $50$ numerical simulations for each point. Simulations were run on a line with $128$ cells with periodic boundary conditions. In all cells the initial depths of soil and surface water were given by the values in the desert state, Eq. (\ref{e:dfp}), while the initial biomass was $\rho_0 = 10$ g m$^{-2}$ in half of the cells ($f=1/2$) and zero in the remaining half. The parameter $\mu$ took three different values for each of the three curves shown: $1$ g m$^{-2}$ (red squares), $~5$ g m$^{-2}$ (green circles) and $~10$ g m$^{-2}$ (blue triangles). Note that for the same parameters and initial conditions the probability of extinction of the deterministic model is essentially one over the whole regime investigated (not shown). See Table \ref{t} for the remaining parameter values.}\label{f:probvsr}
\efi

\clearpage


\begin{table*}
\begin{tabular}{lllr}
\hline
Parameter & Description & Units & Value\\
\hline
$a$ & maximum infiltration rate & d$^{-1}$ &  0.2\\
$b$ & maximum specific water uptake & mm g$^{-1}$ m$^2$ d$^{-1}$ & 0.05\\
$c$ & conversion of water uptake to plants & g mm$^{-1}$ m$^{-2}$& 10\\
$d$ & plant mortality rate & d$^{-1}$ & 0.25\\
$r$ & water loss due to drainage and evaporation & d$^{-1}$ & 0.2\\
$h$ & length of the side of a cell & m & $2$\\
$k_1$ & half-saturation constant of water uptake & mm & 5\\
$k_2$ & half-saturation constant of water infiltration & g m$^{-1}$ &  5\\
$\mu$ & mean contribution of a plant to biomass density &g m$^{-2}$ & 0.1-10\\
$D_\omega$ & diffusion coefficient for soil water & m$^2$ d$^{-1}$ & 0.1\\
$D_\sigma$ & diffusion coefficient for surface water & m$^2$ d$^{-1}$ & 100\\
$K$ & probability of a seed moving to a neighbouring cell & -- & 0.02\\
$L$ & number of cells & -- & 128, $64\times 64$\\
$R$ & rainfall rate & mm d$^{-1}$ & 0.35-0.60\\
$W_0$ &water infiltration rate in the absence of plants & --  & 0.1\\
\hline
\end{tabular}
\caption{Parameter values for the models studied in this paper.}\label{t}
\end{table*}

\end{document}